\documentclass[10pt,onecolumn]{article}

\usepackage{cite}
\usepackage{setspace}
\usepackage[margin=1in]{geometry}
\usepackage{graphicx}
\graphicspath{{fig/}}
\usepackage{amsmath}
\usepackage{amssymb}
\usepackage{setspace}
\usepackage{subfigure}
\usepackage[usenames]{color}
\usepackage{url}

\newcommand{\changedText}{black}

\begin{document}

\onehalfspacing

\title{Background Subtraction for Online Calibration of Baseline RSS in RF Sensing Networks}
\author{Andrea~Edelstein and Michael~Rabbat\thanks{A.~Edelstein and M.~Rabbat are with the Department of Electrical and Computer Engineering, McGill University, Montr\'{e}al, Qu\'{e}bec, Canada. E-mail: andrea.edelstein@mail.mcgill.ca, michael.rabbat@mcgill.ca}}
\maketitle

\begin{abstract}
Radio frequency (RF) sensing networks are a class of wireless sensor networks (WSNs) which use RF signals to accomplish tasks such as passive device-free localization and tracking.  The algorithms used for these tasks usually require access to measurements of baseline received signal strength (RSS) on each link.  However, it is often impossible to collect this calibration data (measurements collected during an offline calibration period when the region of interest is empty of targets). We propose adapting background subtraction methods from the field of computer vision to estimate baseline RSS values from measurements taken while the system is online and obstructions may be present.  This is done by forming an analogy between the intensity of a background pixel in an image and the baseline RSS value of a WSN link and then translating the concepts of temporal similarity, spatial similarity and spatial ergodicity which underlie specific background subtraction algorithms to WSNs. Using experimental data, we show that these techniques are capable of estimating baseline RSS values with enough accuracy that RF tomographic tracking can be carried out in a variety of different environments without the need for a calibration period.
\end{abstract}

\section{Introduction}

\emph{Radio frequency} (RF) tomography is a promising field of research which uses information from the strength of received wireless signals to construct images of physical environments~\cite{Patwari2010}.  RF tomography can be used to image static environments, but many of the technology's most interesting applications are designed to investigate how an environment changes over time.  For instance, RF tomographic tracking can be used to track the movements of a person, through walls, without the need for, e.g., GPS units.

In order for RF tomography algorithms to function, they require access to calibration values: a set of signal strength measurements taken when the system is deployed and transmitting, but not officially online.  During this time, the environment must be static and empty of temporary obstructions (i.e., in a ``background'' state).  Some RF tomography algorithms use these measurements directly; others compare them with measurements taken later, inferring how the environment is changing from how these measurements differ.  Unfortunately, calibration values are often difficult to obtain, since it may be impossible to keep the environment static and empty long enough to collect them.  {\color{\changedText}Additionally, if the environment changes over time, it can be difficult to clear the area and recalibrate the system.}  Little research has addressed whether it is possible to estimate baseline RSS values without access to a calibration period.

In this paper we adapt background subtraction techniques from the field of computer vision to estimate RSS values on wireless links when no calibration period is available, using only measurements taken when the system is already online.

\subsection{Motivation}
The wireless signal measurements required for RF tomography are typically collected using wireless sensor networks (WSNs).  These networks consist of many sensor nodes which are distributed around a region of interest. 
RF tomography applications recover important information not from any external sensor, but by using the nodes' own transceiver to measure the received signal strength (RSS) of the messages sent along wireless links between the nodes.

Knowledge of physical laws governing how obstructions affect RSS is then exploited by RF tomographic imaging to build 3D representations of the surrounding area, with radio waves being used analogously to the X-rays employed in CT scans~\cite{Wilson2010}. By comparing how RSS measurements on different links change over time, RF tomographic tracking follows the movements of people or objects in the region of interest~\cite{Wilson2010Thesis, Wilson2011, Li2011}. Although not specifically under the umbrella of RF tomography, many node localization methods also use RSS measurements~\cite{Patwari2005}.

\subsection{Problem Description}

{\color{\changedText}Because of the small size and relatively low cost of the nodes used in RF sensing networks, these networks are uniquely suited to be deployed on-the-fly~\cite{Wilson2011}. For instance, they would be well-suited to supplant current tracking technology which may depend on GPS, RFID or video technology.  While these systems can be accurate on the range of metres to millimetres~\cite{Gu2009}, they exhibit major drawbacks: they are expensive and they require that tracking hardware be pre-installed.  This is not the case for RF-sensing networks.  Unfortunately, existing algorithms for RF sensing networks cannot currently work in the on-the-fly manner they are best-suited for because they require access to calibration data which must be collected before the system is brought online, during a time when no temporary obstructions are present in the network.  In practical applications, this is often unacceptable.}

In order for RF sensing networks to achieve their true potential to be used on-the-fly, we would like to find a way to approximate the baseline RSS values seen when no temporary obstructions are present.

One way to approximate this value for a given link might be to average many measurements for that link over time, with some no doubt taken while the RSS value is perturbed by the presence of an obstruction but (hopefully) many more taken while the RSS is closer to its baseline value.  Over time, we might hope that the latter values would overwhelm the former, but this is not guaranteed. It would be better if we could approximate the baseline RSS for a given link by averaging \textit{only} over measurements taken on that link at times when it is not obstructed.  

\subsection{Paper Contribution and Organization}

{\color{\changedText}This paper proposes and evaluates methods for online calibration of RF sensing networks. First, we consider a simple algorithm to estimate links' baseline RSS using the mean RSS values on each link over time.  To improve this method's performance, we adopt background subtraction algorithms---designed to determine which pixels in a frame of video are currently static, background pixels---from the field of computer vision ~\cite{Piccardi2004}.  We extend these algorithms to determine which links in a WSN are currently affected by obstructions (in the ``foreground'') and which are not (in the ``background'').  By taking the mean of the measurements seen on a link only at times when it is in the background, we can achieve a better estimate of the link's baseline RSS without needing an offline calibration period.}
We use experimental data to evaluate the performance of the proposed techniques.

Section~\ref{chapter:Ch2} reviews two (of many) RF sensing network applications to establish the importance of estimating baseline RSS.  Section~\ref{chapter:Ch3} gives an overview of background subtraction in the context of image processing, focusing on three particular background subtraction algorithms.  Section~\ref{chapter:Ch4} explains the modifications which must be made to the aforementioned algorithms to adapt them to wireless links instead of pixels.  Section~\ref{chapter:Ch5} presents a comparison of the performance of these background subtraction algorithms using experimental data.  Section~\ref{chapter:Ch6} summarizes our work and discusses future research.

\section{RF Sensing Networks} \label{chapter:Ch2}

In general, the RSS measurement vector $\mathbf{R}(k)$ (in dBm) at a time $k$, which contains one entry for each link, can be decomposed into two parts,
\begin{equation}
\label{eq:WhatIsRSS}
\mathbf{R}^{(k)} = \mathbf{R_B} - \mathbf{R_F}^{(k)}
\end{equation}
where $\mathbf{R_B}$ is the baseline, ``background'' components of the RSS signal which are static over the short term (although they may change over longer timescales, e.g., if the nodes shift due to wind or if their batteries wear down), and $\mathbf{R_F}^{(k)}$ is the ``foreground'' components which are time-varying due, e.g., to people moving through the environment~\cite{Wilson2010}. If the network is empty of temporary obstructions, $\mathbf{R_F}^{(k)}$ equals~$\mathbf{0}$.

When a target is present, the relationship between the location of the target and its effect on $\mathbf{R_F}^{(k)}$ becomes very complicated.  In general, there is no simple mathematical way to recover $\mathbf{R_B}$ or $\mathbf{R_F}^{(k)}$ from $\mathbf{R}^{(k)}$. Most practical applications of RF sensing networks require access either to $\mathbf{R_B}$ or to $\mathbf{R_F}^{(k)}$ but only $\mathbf{R}^{(k)}$ can be measured directly.

\subsection{RF Tomographic Imaging and Tracking} \label{sec:RSSImaging}

Wilson and Patwari's RF tomographic imaging algorithm~\cite{Wilson2011} periodically inspects $\mathbf{R_F}^{(k)}$ to locate temporary obstructions such as people.  This is done using an array of nodes placed around a region of interest inside of which objects will be moving. After the nodes are set in place, they are allowed to transmit amongst themselves for several minutes while no temporary obstructions are present (i.e., while $\mathbf{R_F}^{(k)} =\mathbf{0}$) to estimate $\mathbf{R_B}$.  Obstructions are then allowed to enter the region of interest, while the nodes continue to probe $\mathbf{R}^{(k)}$, the links' RSS values.  The measured value of $\mathbf{R_B}$ is subtracted from $\mathbf{R}^{(k)}$ to determine $\mathbf{R_F}^{(k)}$.  Wilson and Patwari assume that, for a given link $\ell$, any non-zero value of $R_F^{(k)}[\ell]$ is attributable to the presence of an obstruction located somewhere along the direct line-of-sight (LOS) path of link $\ell$.  Thus, once $\mathbf{R_F}^{(k)}$ is known, a maximum a posteriori formulation can be used to determine the most likely obstruction location(s) given the observed $\mathbf{R_F}^{(k)}$. 

Li et al.'s RF tomographic tracking method~\cite{Li2011} operates on a similar principle.  Here, successive values of $\mathbf{R_F}^{(k)}$ are fed to a particle filter which---given a model of the obstruction's movement---uses knowledge of the obstruction's prior positions to track it. 

Without a calibration period, though, and hence with no way to know the value of $\mathbf{R_B}$, neither algorithm specifies how to isolate $\mathbf{R_F}^{(k)}$ from $\mathbf{R}^{(k)}$. This therefore motivates our need to estimate $\mathbf{R_B}$.

\section{Background Subtraction} \label{chapter:Ch3}

Background subtraction originates from the fields of computer vision and image processing.  Using video from a stationary camera, the goal of background subtraction is to differentiate moving ``foreground'' objects from the relatively static background.

In its simplest form, background subtraction evaluates a test at each pixel of each frame of a video
\begin{equation}
\label{eq:SimpleBackgroundSubtraction}
\left|I^{(k)}[n] - B[n]\right|\underset{\mathcal{B}}{\overset{\mathcal{F}}{\lessgtr}}\theta
\end{equation}
where $I^{(k)}[n]$ is the value of $n$th pixel in frame $k$ (e.g., intensity on a scale from~0 to~255 for greyscale video), $B[n]$ is the known or estimated value of the true background at $n$ and $\theta$ is a threshold parameter~\cite{McHugh2008}.  Evaluating this test assigns $n$ to the foreground set $\mathcal{F}$ or to the background set $\mathcal{B}$.

In practice, background subtraction algorithms must account for several factors including: a background which is not completely static (e.g., due to fluttering leaves), noise in the image (e.g., due to camera jitter) and changes to the background over time (e.g., as the sun sets).  Using a non-zero threshold in (\ref{eq:SimpleBackgroundSubtraction}) can compensate for some of these problems, but determining the correct model for $B$ and the correct value for $\theta$ can be difficult.  Many background subtraction algorithms have been developed to cope with these issues.  We now describe three such methods.

\subsection[Temporal Background Modelling]{Temporal Background Modelling}\label{sec:TBM}

The most basic method we consider seeks only to model the distribution of the background by using a nonparametric Gaussian kernel density function~\cite{Elgammal2002} and by relying on the concept of temporal similarity, the assumption that if a pixel is in the background, measurements taken on it within a short time frame should be similar to one another.  Note that this temporal similarity is not expected to hold over long periods of time; e.g., the background may evolve over long time-scales.  More formally, this model dictates that the background PDF $P_\mathcal{B}$ of the intensity $I^{(k)}[n]$ of a pixel $n$ of a frame $k$ of video can be modelled as
\begin{equation}
\label{eq:BackgroundPDF}
P_\mathcal{B}\left(I^{(k)}[n]\right) = \frac{1}{N}\sum_{i=1}^{N}\mathcal{K}\left(I^{(k)}[n]-I^{(k-i)}[n]\right)
\end{equation}
where $\mathcal{K}(\cdot)$ is a zero-mean Gaussian kernel function with a variance $\sigma_t^2$ which is assumed to be constant.  Here, $N$ refers to the number of recent frames used to model the background at frame $k$.

Once $P_\mathcal{B}$ has been formed in this way, a simple test
\begin{equation}
\label{eq:SimpleTest}
\frac{P_\mathcal{B}\left(I^{(k)}[n]\right)}{P_\mathcal{F}\left(I^{(k)}[n]\right)}\underset{\mathcal{B}}{\overset{\mathcal{F}}{\lessgtr}}\eta\frac{\pi_\mathcal{F}}{\pi_\mathcal{B}}
\end{equation}
is carried out to determine if each pixel in each frame is in the background.  In this equation, $\eta$ is a cost term which can be used to penalize or favour different classification errors, $\pi_\mathcal{F}$ and $\pi_\mathcal{B}$ are the prior probabilities that a pixel is in the background or the foreground and $P_\mathcal{F}$ is the foreground PDF which is assumed to be constant over $n$ and $k$; this assumption will be modified in Sections~\ref{sec:FABS} and~\ref{sec:FABSMMCL}.  Likewise, for now, $\pi_\mathcal{F}$ and $\pi_\mathcal{B}$ are considered to be constant and equal.  The constants in (\ref{eq:SimpleTest}) can be combined into a single new constant $\theta$, turning the equation into
\begin{equation}
\label{eq:SimpleTestTheta}
P_\mathcal{B}\left(I^{(k)}[n]\right)\underset{\mathcal{B}}{\overset{\mathcal{F}}{\lessgtr}}\theta.
\end{equation}

This approach is referred to as background subtraction with \emph{temporal background modelling} (TBM).  Compared to the other background subtraction algorithms described below, TBM has lower computational complexity and fewer parameters which need to be tuned ($N$, $\sigma_t^2$ and $\theta$). However, more sophisticated methods often yield superior performance.

\subsection{Foreground-Adaptive Background Subtraction}\label{sec:FABS}

\emph{Foreground-adaptive background subtraction} (FABS)~\cite{McHugh2008,McHugh2009} uses the methods described in Section~\ref{sec:TBM} to model $P_\mathcal{B}$.  It then models $P_\mathcal{F}$ using spatial similarity, the idea that foreground pixels in the same image which are close together will tend to be similar in value. More specifically, McHugh et al.~\cite{McHugh2009} employ the concept of a neighbourhood $\mathcal{N}(n)$: a square of a certain size centered on $n$, inside which all foreground pixels will tend to be similar. For instance, say we know that pixel $n$ is part of the face of a person in the foreground.  Any foreground pixels in the neighbourhood of $n$ will then likely be flesh-coloured as well. Thus, spatial similarity is the assumption that foreground objects occupy space (a group of adjacent pixels) and are spatially homogenous in value.

To model the foreground, the TBM test (\ref{eq:SimpleTestTheta}) is first applied to a given frame to assign a preliminary label to each pixel.  Then, an initial model of the foreground, $P_{\mathcal{F}_0}$, is built according to
\begin{equation}
\label{eq:ForegroundPDF}
P_{\mathcal{F}_0}\left(I^{(k)}[n]\right) = \frac{\sum_{m\in\mathcal{N}_{\mathcal{F}_0}^{(k)}(n)}\mathcal{K}\left(I^{(k)}[n]-I^{(k)}[m]\right)}{\left|\mathcal{N}_{\mathcal{F}_0}^{(k)}(n)\right|}
\end{equation}
where $\mathcal{N}_{\mathcal{F}_0}(n)$ is a set containing all the neighbours of $n$ which have been assigned a foreground label by the preliminary TBM test.  Again, $\mathcal{K}(\cdot)$ is a zero-mean Gaussian kernel function, this time with constant variance $\sigma_s^2$.  Once $P_{\mathcal{F}_0}$ is calculated, it is used in
\begin{equation}
\label{eq:ForegroundTest}
\frac{P_\mathcal{B}\left(I^{(k)}[n]\right)}{P_{\mathcal{F}_0}\left(I^{(k)}[n]\right)}\underset{\mathcal{B}}{\overset{\mathcal{F}}{\lessgtr}}\eta
\end{equation}
to assign new labels to each pixel.  These labels are then used to build $\mathcal{N}_{\mathcal{F}_1}$, $P_{\mathcal{F}_1}$, $\mathcal{N}_{\mathcal{F}_2}$, $P_{\mathcal{F}_2}$, ... until a convergence criterion is satisfied.  McHugh \cite{McHugh2008} suggests that only 2--3 iterations of this process are usually needed for the labels to stabilize to what can be called $P_\mathcal{F}$ (with no iterative subscript).
 
While the sets of background and foreground pixels returned by FABS should be closer to the ground truth than those returned by TBM, note that FABS introduces positive feedback to the background subtraction process.  FABS is specifically designed to look for foreground pixels which have erroneously been left in the background by a simpler background subtraction algorithm.  These pixels are then added to $\mathcal{F}$.  This means that if $\eta$ is set too high, FABS may eventually place every pixel in the image in $\mathcal{F}$.  The threshold $\eta$ must be chosen carefully to avoid this.

We point out here that FABS is more computationally-complex than TBM.  It also introduces several more parameters which must be set, namely $\eta$, $\sigma_s^2$ and neighbourhood size/structure. 

\subsection[Markov Modelling of Change Labels]{Foreground-Adaptive Background Subtraction with Markov Modelling of Change Labels}\label{sec:FABSMMCL}

\emph{Foreground-adaptive background subtraction with Markov modelling of change labels} (FABS-MMCL)~\cite{McHugh2008,McHugh2009} employs the aforementioned methods to model $P_\mathcal{B}$ and $P_\mathcal{F}$ and then models $\pi_\mathcal{F}$ and $\pi_\mathcal{B}$ by relying on spatial ergodicity, the idea that pixels which are close together will tend to belong either to the foreground or the background together.

{\color{\changedText}McHugh et al.~\cite{McHugh2009} propose that the FABS test in Equation~\eqref{eq:ForegroundTest} first be used to label each pixel in a given frame.  McHugh~\cite{McHugh2008} then points out that, by assuming the pixels' true labels obey the Markov property, these labels represent a realization of a Markov random field $G$.  The a priori probabilities of such a field taking a certain value follow the Gibbs distribution
\begin{equation}
\label{eq:GibbsDistribution}
P(G = g) = \frac{1}{Z}\exp\left(\frac{-1}{\gamma}\sum_{\left\{n,n'\right\}\in\mathcal{Q}}V\left(n,n'\right)\right)
\end{equation}
where $Z$ is a normalization constant and $\gamma$ is the distribution's natural temperature.  $\mathcal{Q}$ corresponds to a pixel neighbourhood, and $V\left(\cdot\right)$ is an indicator function which equals~0 if pixels $n$ and $n'$ have the same label and~1 if they do not.  This prior penalizes pixels which have labels that are very different from their neighbors (i.e., showing a high degree of non-ergodicity), and discourages isolated false positives.

By temporarily assuming that $G$ is known for all pixels except $n$,~\eqref{eq:GibbsDistribution} can be combined with~\eqref{eq:SimpleTest} to write
\begin{equation}
\label{eq:MarkovTest}
\frac{P_\mathcal{B}\left(I^{(k)}[n]\right)}{P_\mathcal{F}\left(I^{(k)}[n]\right)}\underset{\mathcal{B}}{\overset{\mathcal{F}}{\lessgtr}}\eta\exp\left(\frac{1}{\gamma}\left(\left|\mathcal{N}^{(k)}_\mathcal{F}(n)\right| - \left|\mathcal{N}^{(k)}_\mathcal{B}(n)\right|\right)\right).
\end{equation}
Here $\left|\mathcal{N}_\mathcal{F}(n)\right|$ and $\left|\mathcal{N}_\mathcal{B}(n)\right|$ are the number of foreground and background neighbours of $n$ as labelled by the first application of the FABS test in (\ref{eq:ForegroundTest}). Labels are then updated via the FABS-MMCL test~(\ref{eq:MarkovTest}).

McHugh et al.~\cite{McHugh2009} recommend repeatedly applying the test \eqref{eq:MarkovTest} to refine the labels (a technique known as iterated conditional modes~\cite{Besag1986}). Although the recommended number of iterations is relatively low ($\approx$~10), FABS-MMCL is the most computationally-intensive of the three methods and has the most parameters to set.}

Note that FABS-MMCL is designed in part to prune back the effects of the overly-greedy FABS algorithm.  Because of this, the values of $\eta$, $\sigma_s^2$ and the neighbourhood size which produce the best results for FABS in isolation may not be optimal when FABS outputs are used as input to FABS-MMCL.

\section{Adaptation of Background Subtraction to WSNs}\label{chapter:Ch4}

{\color{\changedText}Before adapting background subtraction to WSNs, we ask whether any simpler techniques can be used to determine $\mathbf{R_B}$ when it cannot be measured directly.

One naive approach would be simply to average over all available RSS measurements for a given link.  However, if an obstruction is moving around or across the link, this average will incorporate the obstruction's effects.  We will still introduce this basic \emph{mean approximation} (MA) approach as a starting point since it is easy to implement and nonetheless an improvement over the current situation wherein no ways have yet been proposed to determine $\mathbf{R_B}$ at all.}

Another naive approach would examine the RSS for a given link over a window of time, searching for periods when the RSS drops suddenly.  One might then attribute these drops to the attenuation caused by a newly-appearing obstruction, use a threshold to discard these values and take the mean of the remaining measurements.  Unfortunately, the effect of obstructions on RSS is not that straightforward; due to multi-path effects which can cause constructive interference, obstructions can cause RSS to decrease \textit{or} increase.  Decreasing battery voltage and shifts in the nodes' positions will also cause changes to the RSS, but we do not wish to discard measurements which are perturbed for these reasons.  Finally, a considerable amount of noise is seen even on unobstructed links~\cite{Edelstein2011}.  These factors drastically reduce the performance of simple techniques such as thresholding~\cite{Moussa2009}.

\subsection{Basic Background Subtraction in WSNs}\label{sec:BasicWSNBackgroundSubtraction}

The naive algorithms mentioned above are not always equipped to deal with the complex behaviour of wireless signals.  However, we can see strong parallels between the problem of determining which links in a network are currently affected by a moving obstruction (i.e., are in the foreground) and that of determining which pixels in an image are currently displaying a moving foreground object, pointing to the utility of background subtraction in the WSN domain. Before background subtraction can be applied to a WSN, it is necessary to establish what constitutes a ``pixel,'' a measurement of ``intensity'' and a ``frame'' in this context.  We can make the association between a single pixel $n$ in an image and a single bidirectional link $\ell$ in a WSN, and we can see that the discretized value of the RSS (in dBm) measured on a link during an as-yet-to-be-defined frame $k$, $R^{(k)}[\ell]$, represents an analogue to the greyscale intensity of a pixel, $I^{(k)}[n]$.

In the image processing domains, one frame of video comprises one measurement for each pixel in the video image.  Ideally, we would like our corresponding RSS frame to contain one RSS measurement for each bidirectional link with all these measurements taken at exactly the same time, giving us an instantaneous capture of the RSS values on all the links in the network.  Because WSNs are subject to interference and packet collisions when multiple nodes transmit at the same time, this is not realistic. 

In practice, RSS measurements are usually obtained as follows: for a network with $M$ nodes, node~1 broadcasts a packet to all other nodes.  This establishes measurements for the RSS on $M - $1 unidirectional links stretching from node~1 to node~2, from node~1 to node~3, etc.  Then, node~2 broadcasts a packet to all the other nodes, yielding another series of $M - $1 unidirectional RSS measurements.  This continues until all $M$ nodes have broadcast, at which point node~1 broadcasts again.  Therefore, it is impossible to take simultaneous measurements on all the links in the network.  Fortunately, WSNs are capable of taking many such measurements per second: in our experiments, a 22-node network could complete such a cycle in $\approx$~120~ms.  Because this time span is so short, we assume the conditions in the network (i.e., obstruction locations) do not change in the time it takes to complete one measurement cycle.

We must also address that, because WSNs are subject to dropped packets, a single measurement cycle may not return a value for every link.  If this happens, we do not want to wait for additional cycles to complete our frame.  Thus, instead of a frame being comprised of one measurement for each link, we define a frame as being built from $M$ consecutive measurement vectors, where $M$ is the number of nodes in the network. Ideally, each node contributes one measurement vector to each frame. However, if the message from one node is dropped then that node may be omitted and another node included multiple times. In this case, we average the multiple measurement vectors for the node which appears in the same window twice. Then, the two directed measurements for each transceiver pair are averaged together to create one measurement for each uni-directional link.  Finally, if measurements have been dropped for one or more links, these values are copied from the previous frame.  The next frame is then created from the next $M$ consecutive measurement vectors and so on.  A consequence of this method is that there is no overlap in the measurements used from frame to frame, except where dropped packets force us to reuse values.

Having established these analogies for pixels, pixel intensities and frames, we see that the temporal similarity underlying TBM should still hold here to some degree: if the effects of noise are small, the consecutive measurements taken on a given link should not vary while that link stays in the background (discounting long-term changes such as those caused by a dying battery).  In that case, we can calculate the background PDF $P_\mathcal{B}$ via an updated version of (\ref{eq:BackgroundPDF}) such that
\begin{equation}
\label{eq:BackgroundPDFWSN}
P_\mathcal{B}\left(R^{(k)}[\ell]\right) = \frac{1}{N}\sum_{i=1}^{N}\mathcal{K}\left(R^{(k)}[\ell]-R^{(k-i)}[\ell]\right)
\end{equation}
and we can build our sets of background and foreground links via an updated version of (\ref{eq:SimpleTestTheta}) such that
\begin{equation}
\label{eq:SimpleTestThetaWSN}
P_\mathcal{B}\left(R^{(k)}[\ell]\right)\underset{\mathcal{B}}{\overset{\mathcal{F}}{\lessgtr}}\theta.
\end{equation}
Recall that $N$ is the number of recent frames used to model the background at frame $k$ and $R^{(k)}[\ell]$ is the measured RSS value for link $\ell$ at frame $k$.

\subsection{Spatial Similarity in WSNs}\label{sec:SpatialSimilarityInWSNs}

While determining analogues for pixels, pixel intensity and frames is relatively straight-forward, applying spatial similarity and the concept of neighbourhoods to WSNs is more complicated: there is no reason why two foreground links which are close together must have similar RSS values.  Consider a network of nodes deployed around the perimeter of a square.  Let $\ell_1$ be the link connecting the nodes at $(0,0)$ and $(0,1)$, let $\ell_2$ be the link connecting the nodes at $(0,0)$ and $(0,7)$ and let both these links be obstructed by the same object.  The two overlapping links are physically close together, but because $\ell_1$ is 1~m long while $\ell_2$ is 7~m long, the log-distance path loss model suggests that their RSS values will be quite different.

Instead, to capture the notion of similarity between links, we define the neighbourhood $\mathcal{L}(\ell)$ of link $\ell$ to be the set of links similar in length to $\ell$.  Letting $d_\ell$ represent the length of link $\ell$ (i.e., distance between transmitter and receiver), $\mathcal{L}(\ell)$ is thus defined as
\begin{equation}
\label{eq:LNeighbourhoodDefinition}
\mathcal{L}(\ell) = \{\ell': |d_\ell - d_{\ell'}| \le \tau\},
\end{equation}
where $\tau \geq 0$ is a parameter to be tuned. This definition implies the creation of $\mathcal{L}_{\mathcal{F}}^{(k)}(\ell)$, the set of foreground neighbours associated with $\mathcal{L}(\ell)$ in frame $k$. In this way, spatial similarity in the image domain becomes similarity-of-length in the WSN domain.

When formulating $\mathcal{L}(\ell)$ in this way, we are aware that not all links in this type of neighbourhood are guaranteed to have similar RSS values.  Even for links of equal length, if one~1~m link in the network passes through a wall while another passes through free space, their values will be different.  Keep in mind, though, that even in the image processing domain, not all foreground pixels in a neighbourhood are guaranteed to have a similar intensity (e.g., a flesh-coloured pixel may be next to a red pixel from a shirt).

Furthermore, we are aware that two links of the same length whose LOS paths are both obstructed by the same moving object may have different RSS values if the object intersects them in different ways, e.g., if an obstruction is located at the end of one link and in the center of the other~\cite{Rappaport2002}.  In practice, though, we find that this difference is within the range of the normal variations due to noise seen on the links.  

We can now present new expressions for the model of the preliminary foreground PDF $P_{\mathcal{F}_0}$ (originally in (\ref{eq:ForegroundPDF})) and for the FABS test (originally in (\ref{eq:ForegroundTest})) such that
\begin{equation}
\label{eq:ForegroundPDFWSN}
P_{\mathcal{F}_0}\left(R^{(k)}[\ell]\right) = \frac{\sum_{m\in\mathcal{L}_{\mathcal{F}_0}^{(k)}(\ell)}\mathcal{K}\left(R^{(k)}[\ell]-R^{(k)}[m]\right)}{\left|\mathcal{L}_{\mathcal{F}_0}^{(k)}(\ell)\right|}
\end{equation}
and 
\begin{equation}
\label{eq:ForegroundTestWSN}
\frac{P_\mathcal{B}\left(R^{(k)}[\ell]\right)}{P_\mathcal{F}\left(R^{(k)}[\ell]\right)}\underset{\mathcal{B}}{\overset{\mathcal{F}}{\lessgtr}}\eta.
\end{equation}

\subsection{Spatial Ergodicity in WSNs}\label{sec:SpatialErgodicityInWSNs}

We also consider the concept of spatial ergodicity which underlies FABS-MMCL.  Spatial ergodicity carries over to WSNs more directly than does spatial similarity since we still say that if two links are ``close together'', they are likely to pass through or avoid the same obstructions and hence have correlated foreground or background states.  Nonetheless, we need to reconsider what it means for two links to be ``close together.''  Must they share a common endpoint?  Must their LOS paths overlap or intersect? The following subsections describe two methods for determining whether two links are close enough together for spatial ergodicity to hold.  

\subsubsection{Rectangle-Based Neighbourhoods} \label{sec:RectangleBasedMethod}

\begin{figure}
\centering
\includegraphics[width=2.4in]{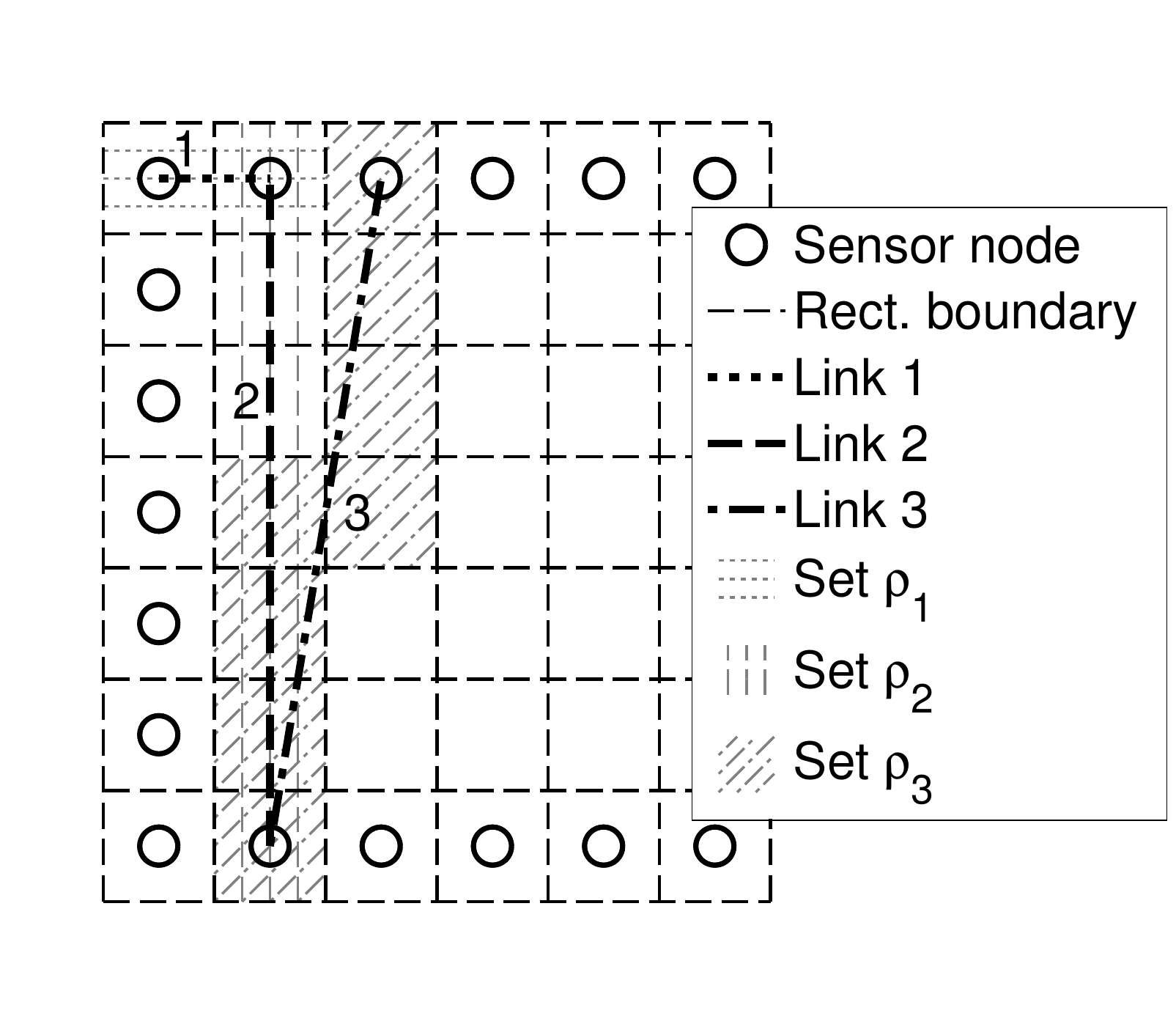}
\caption{{\color{\changedText}A graphical representation of FABS-MMCL-R.  The set of rectangles $\rho$ associated with each link is shaded.  We see $\left|\rho_1\right|=2$, $\left|\rho_2\right| = 7$ and $\left|\rho_1\cap\rho_2\right|=1$, creating shared rectangle percentage ${\chi\left(\ell_1,\ell_2\right)=\frac{1}{\min\left(2,7\right)}=0.5}$.  Similarly, $\left|\rho_3\right| = 8$, $\left|\rho_2\cap\rho_3\right|=4$ and ${\chi\left(\ell_2,\ell_3\right)=\frac{4}{\min\left(7,8\right)}=0.57}$.}}
\label{fig:RectanglesDiagram}
\end{figure}

The first method involves partitioning the interior of the region of interest into rectangles whose side lengths are determined by the node separations.  These rectangles are similar to the cubic voxels proposed by Wilson and Patwari in~\cite{Wilson2011}.  We denote by $\rho$ the set of rectangles through which a link $\ell$ passes.  Then, for two links, $\ell_1$ and $\ell_2$ with associated sets of rectangles $\rho_1$ and $\rho_2$, we define $\chi\left(\ell_1,\ell_2\right)$, the percentage of rectangles common to both links as
\begin{equation}
\label{eq:CommonRectangles}
\chi\left(\ell_1,\ell_2\right) = \frac{\left|\rho_1\cap\rho_2\right|}{\min\left(\left|\rho_1\right|,\left|\rho_2\right|\right)} \times 100.
\end{equation}
If this percentage is larger than a certain tuneable threshold, we deem $\ell_1$ and $\ell_2$ to be neighbours.  A graphical representation of this process can be seen in Fig.~\ref{fig:RectanglesDiagram}. 
We refer to this method as FABS-MMCL-R.

Unlike in the computer vision and image processing domains, the WSN-domain neighbourhoods used for FABS and FABS-MMCL are not the same sets.  As previously mentioned, we denote the neighbourhood for FABS as $\mathcal{L}(\ell)$.  We now denote the neighbourhood used for FABS-MMCL as $\mathcal{S}(\ell)$.  For FABS-MMCL-R, this neighbourhood is defined as
\begin{equation}
\label{eq:SNeighbourhoodDefinition}
\mathcal{S}(\ell) = \{\ell' : \chi(\ell, \ell') \geq C\}
\end{equation}
where $0<C\leq100$ is left as a parameter to be tuned.  We also define $\mathcal{S}_{\mathcal{F}}^{(k)}(\ell)$ and $\mathcal{S}_{\mathcal{B}}^{(k)}(\ell)$, the sets of foreground and background neighbours, respectively, associated with $\mathcal{S}(\ell)$ in frame $k$. This leads to the FABS-MMCL test (\ref{eq:MarkovTest}) becoming
\begin{equation}
\label{eq:MarkovTestWSN}
\frac{P_\mathcal{B}\left(R^{(k)}[\ell]\right)}{P_\mathcal{F}\left(R^{(k)}[\ell]\right)}\underset{\mathcal{B}}{\overset{\mathcal{F}}{\lessgtr}}\eta\exp\left(\frac{1}{\gamma}\left(\left|\mathcal{S}^{(k)}_\mathcal{F}(\ell)\right| - \left|\mathcal{S}^{(k)}_\mathcal{B}(\ell)\right|\right)\right).
\end{equation}

\subsubsection{Obstruction-Based Neighbourhoods} \label{sec:ObstructionBasedMethod}

\begin{figure}
\centering
\includegraphics[width=1.65in]{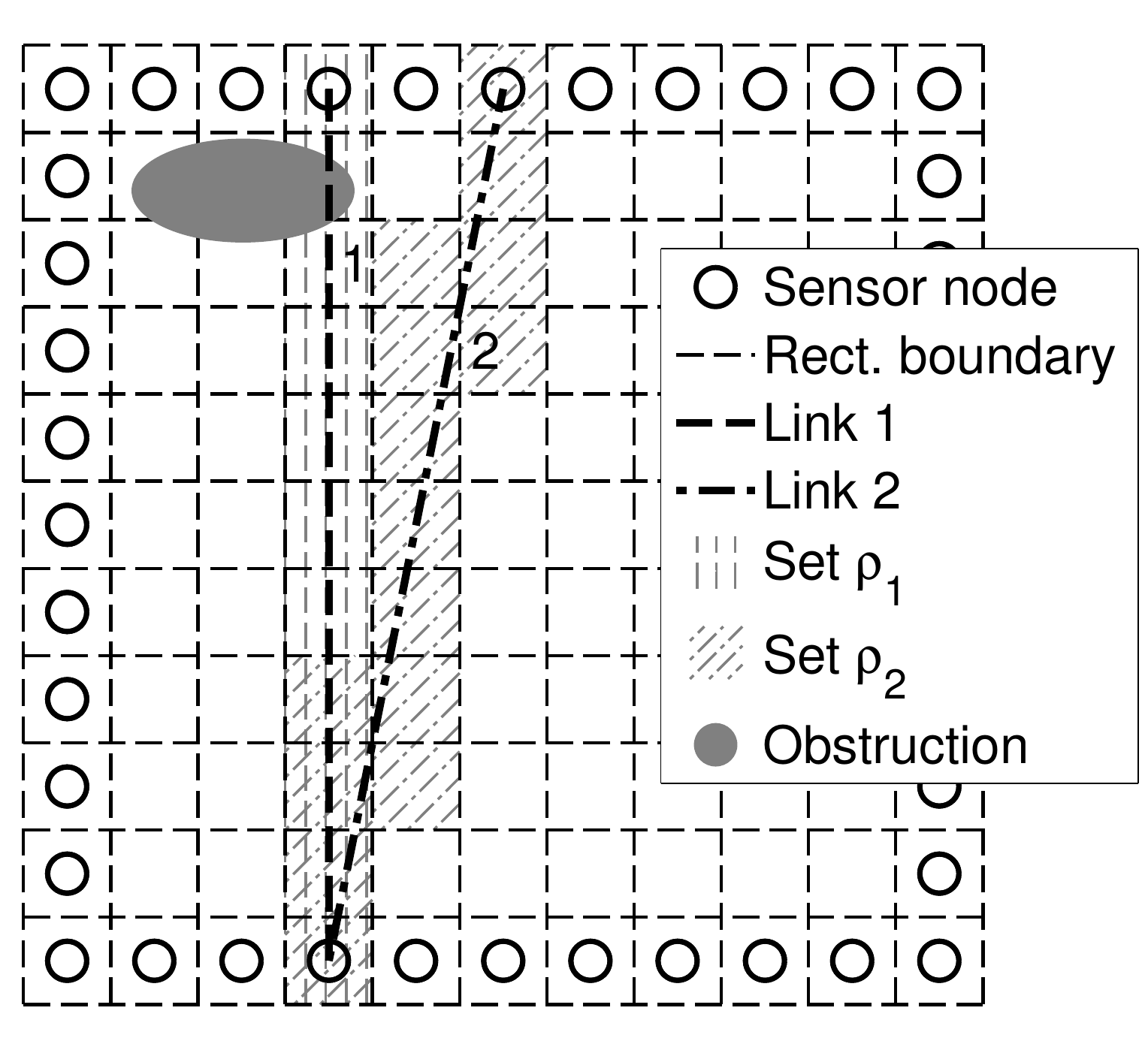}
\caption{{\color{\changedText}A possible relationship between an obstruction and two different links.  Links $\ell_1$ and $\ell_2$ may pass through a large enough percentage of common rectangles to qualify as neighbours. Nonetheless, as seen at the top of the figure, the links' paths may eventually diverge enough that obstructions (like the one pictured) do not affect them both.}}
\label{fig:DivergingLinks}
\end{figure}

{\color{\changedText}This section describes an alternative, data-driven way to define the neighborhood set $S(\ell)$ used by FABS-MMCL. Although FABS-MMCL-R is fairly intuitive, it may have problems.  For instance, if two LOS links have a small angle of separation, they may be close enough to cross the same rectangles over a large fraction of their length, thereby qualifying as neighbours according to~\eqref{eq:SNeighbourhoodDefinition}.  However, this is misleading if their paths diverge significantly by the time one of them crosses the spot where an obstruction is located, as seen in Fig.~\ref{fig:DivergingLinks}. This section discusses an alternative approach to determine neighborhoods based on estimates of where the obstruction may be located.}

\begin{figure}
\centering
\includegraphics[width=1.6in]{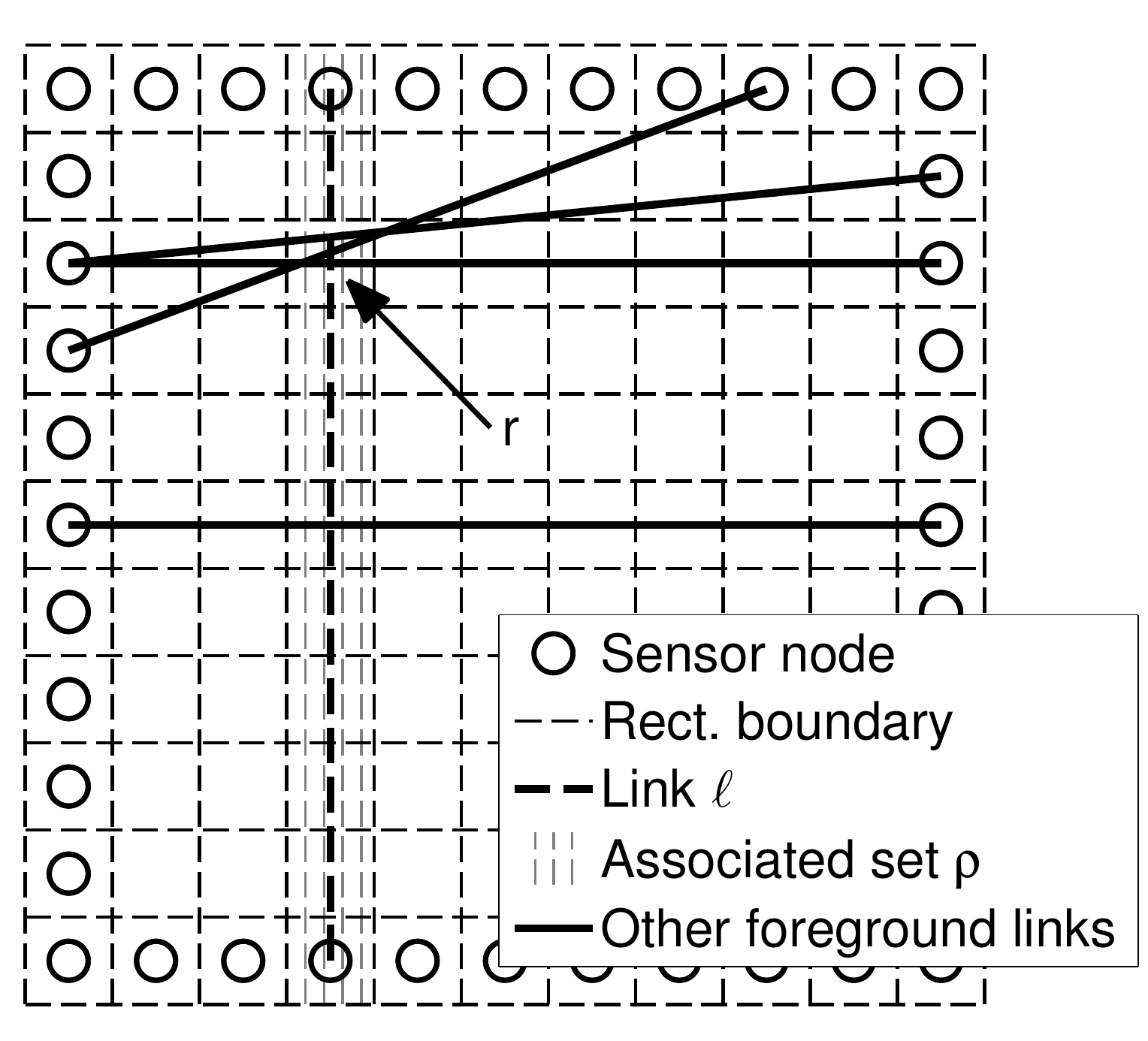}
\caption{{\color{\changedText}A graphical representation of FABS-MMCL-O.  Link $\ell$ is the vertical link.  Other foreground links (as designated by FABS or by previous iterations of FABS-MMCL-O) are also shown.  Excluding $\ell$ itself, one pixel in $\rho$ is crossed by~3 foreground links, one pixel in $\rho$ is crossed by~1 foreground link and the rest of the pixels in $\rho$ are crossed by~0 foreground links.  Since ${\max(3,1,0)=3}$, $r$ is the rectangle crossed by~3 foreground links.}}
\label{fig:FABSMMCLO}
\end{figure}

Recall that we do not have any a priori information about the location of these obstructions.  We can still hypothesize, though, that their locations may be inferred by looking for rectangles which contain particularly large numbers of foreground links.  Specifically, just before each application of (\ref{eq:MarkovTestWSN}), we can count the number of foreground links passing through each rectangle in $\rho$.  We can then find the rectangle $r$ in $\rho$ with the largest number of foreground links passing through it and assign $\mathcal{S}^{(k)}(\ell)$ to be the set of links which pass through $r$.  This is shown graphically in Fig.~\ref{fig:FABSMMCLO}.  We refer to this method as FABS-MMCL-O.

Of course, $r$ may not be located near an obstruction at all since, in fact, it is quite possible that none of the rectangles in $\rho$ are near the obstruction (e.g., if the LOS of $\ell$ is positioned entirely on the extreme east side of the network while the obstruction is on the west side of the network).  If this is the case, even though $r$ may contain more foreground links than any other rectangle in $\rho$, $\left|\mathcal{S}^{(k)}_\mathcal{F}(\ell)\right|$ is still likely to be small.  This will cause the exponent term in the FABS-MMCL test (seen in (\ref{eq:MarkovTestWSN})) to be highly negative, decreasing the right hand side of the equation.  In this way, no harm is done if $r$ is artificially deemed to contain an obstruction.  However, in our experiments with the TBM method, we notice that even when we know for a fact that $r$ contains an obstruction, and even when our classification of foreground and background links results in an accurate estimate of the baseline RSS, there may still be relatively few foreground links and hence many background links passing through $r$.  This means that, if we apply FABS-MMCL-O to the output of the TBM algorithm as described, the exponential term in (\ref{eq:MarkovTestWSN}) would almost always be highly-negative.  To counteract this tendency, we introduce a term $\mu$ to the FABS-MMCL test such that
\begin{equation}
\label{eq:MarkovTestOWSN}
\frac{P_\mathcal{B}\left(R^{(k)}[\ell]\right)}{P_\mathcal{F}\left(R^{(k)}[\ell]\right)}\underset{\mathcal{B}}{\overset{\mathcal{F}}{\lessgtr}}\eta\exp\left(\frac{1}{\gamma}\left(\left|\mathcal{S}^{(k)}_\mathcal{F}(\ell)\right| - \left|\mathcal{S}^{(k)}_\mathcal{B}(\ell)\right|+\mu\right)\right).
\end{equation}
Recall that $\mathcal{B}$ and $\mathcal{F}$ are the sets of background and foreground links respectively.  Then, by resetting \mbox{$\mu = \left|\mathcal{B}\right|-\left|\mathcal{F}\right|$} before each application of (\ref{eq:MarkovTestOWSN}), we ensure that the distribution of the exponent is roughly centered around~0 instead of always being negative.

\subsubsection{Other Methods of Building Neighbourhoods} \label{sec:OtherMethods}

Although we present two models for building $\mathcal{S}(\ell)$, we are aware that other researchers have considered related questions.  In~\cite{Gudmundson1991}, Gudmundson proposes a model for quantifying the correlation of the shadowing observed on different links, but this model only applies to links with at least one common endpoint, which is not the case for most links in a WSN.  Wang et al.~\cite{Wang2006} extend Gudmundson's work to model shadowing correlation between two links, $\ell_1$ and $\ell_2$, when $\ell_2$ is created by moving the nodes which form $\ell_1$ to a new location.  We might apply this model to a case where $\ell_1$ and $\ell_2$ are two links defined by four completely separate senders and receivers, but Wang et al.'s model also assumes that the distance between $\ell_1$ and $\ell_2$ is small compared with the lengths of the links themselves.  Again, this does not apply to most links in a WSN.  {\color{\changedText}Agrawal and Patwari~\cite{Patwari2008,Agrawal2009} also propose the network shadowing (NeSh) model to calculate the shadowing correlation seen across links in a wireless network, and this may also be used as an alternative method to define link neighborhoods.}

\section[Experimental Evaluation]{Experimental Evaluation}\label{chapter:Ch5}

\subsection{Experiment Setup}\label{sec:ExperimentSetup}

Through an association with the Beijing University of Posts and Telecommunications (BUPT) in Beijing, China, we have access to data from a number of RF sensing experiment sets carried out by BUPT students.  Each set of experiments used the same methodology, hardware and software.  Only the number of nodes and the environment in which they were deployed varied between experiment sets.

The specific hardware employed in these experiments consisted of a fleet of Texas Instruments CC2530 System-on-Chips equipped with built-in RSS indicators capable of measuring RSS to a resolution of 1~dBm.  These nodes conform to the IEEE 802.15.4 standard, transmitting in the 2.4~GHz band with a configurable output power~\cite{CC2530}.  For these experiments, the nodes were set to transmit at 0~dBm and each node was augmented with an external antenna.

The nodes were programmed to follow the token-ring protocol from Section~\ref{sec:BasicWSNBackgroundSubtraction}, and all $M$ nodes communicated with one another, forming a fully-connected topology which could directly measure the RSS between every pair of nodes in the network.  The system was configured to probe this RSS (by having the next node in the ring send a new wireless message) approximately every 5~ms.  The nodes also sent their measurements to an $M + $1st node, which forwarded the data via USB cable to a laptop, which recorded the values.  This extra node acted only as an intermediary to the laptop; any RSS values on links involving this node are irrelevant to the experiments.

These experiments were conducted as follows: In each new environment, the nodes were mounted on~$\approx$~1~m-high stands around the perimeter of the region of interest.  They were then allowed to transmit amongst themselves for several minutes while the network was kept vacant, measuring the baseline RSS on each link directly.  After this offline calibration period was complete, a person entered the region of interest and walked around the nodes.  This person followed a path marked on the ground, and a timestamp was recorded as they passed various points on their route so that their location was known at all times.  Each experiment set thus consists of baseline RSS data collected while the region of interest was empty, coupled with data collected during one or more ``walks.''  Note that the length of each walk varied; some lasted only long enough for~$\approx$~150 frames of measurements to be obtained, while in other cases, up to~700 frames of measurements were collected.  

The first set of experiments was conducted in an outdoor field at BUPT.  Outdoor environments such as fields are often the most conducive to RF tomography applications as the lack of walls and furniture means that troublesome multipath effects are far less prominent.  In this case, 22 nodes were placed in a 7~m~$\times$~7~m square, as seen in Fig.~\ref{fig:PlayAgainExperiment1}.  The odd node positioning in the upper-right corner is due to the fact that two nodes were discovered to be defective after the experiments had been run and the measurements taken by these nodes had to be discarded.  The walker's path was a 6~m~$\times$~6~m square inside the nodes.

The second set of experiments was conducted in an indoor lab, with 24 nodes positioned in a 3~m~$\times$~3.6~m rectangle (see Fig.~\ref{fig:PlayAgainExperiment2}).  The walker's path was a 1.8~m~$\times$~2.4~m rectangle inside the nodes. Although there was furniture along the walls of the room (outside the region being monitored by the nodes), no furniture was located inside the region of interest.  Thus, the only thing obstructing the LOS paths of the wireless signals was the walker sent into the region of interest; nonetheless---as in indoor environments in general---the walls, floor, ceiling and furniture still contribute to the multipath effects which generally make RF sensing more challenging indoors.

\begin{figure}
\centering
\subfigure[Photo of the experiment setup (seen from inside the lab).]{
\includegraphics[height = 3.03cm]{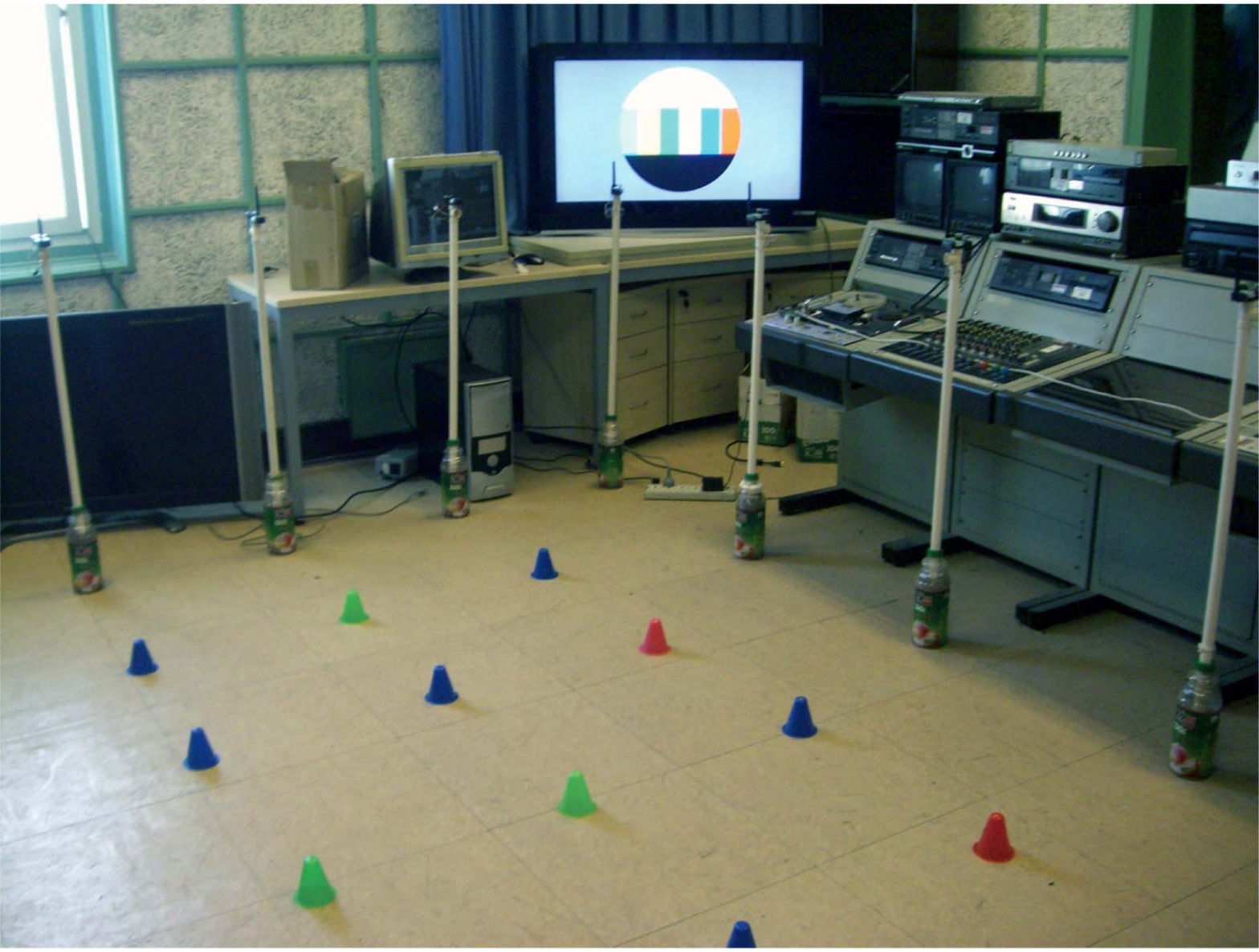}}
\subfigure[Photo of the experiment setup (seen from the hall).  This door was closed for the experiment.]{
\includegraphics[height = 3.03cm]{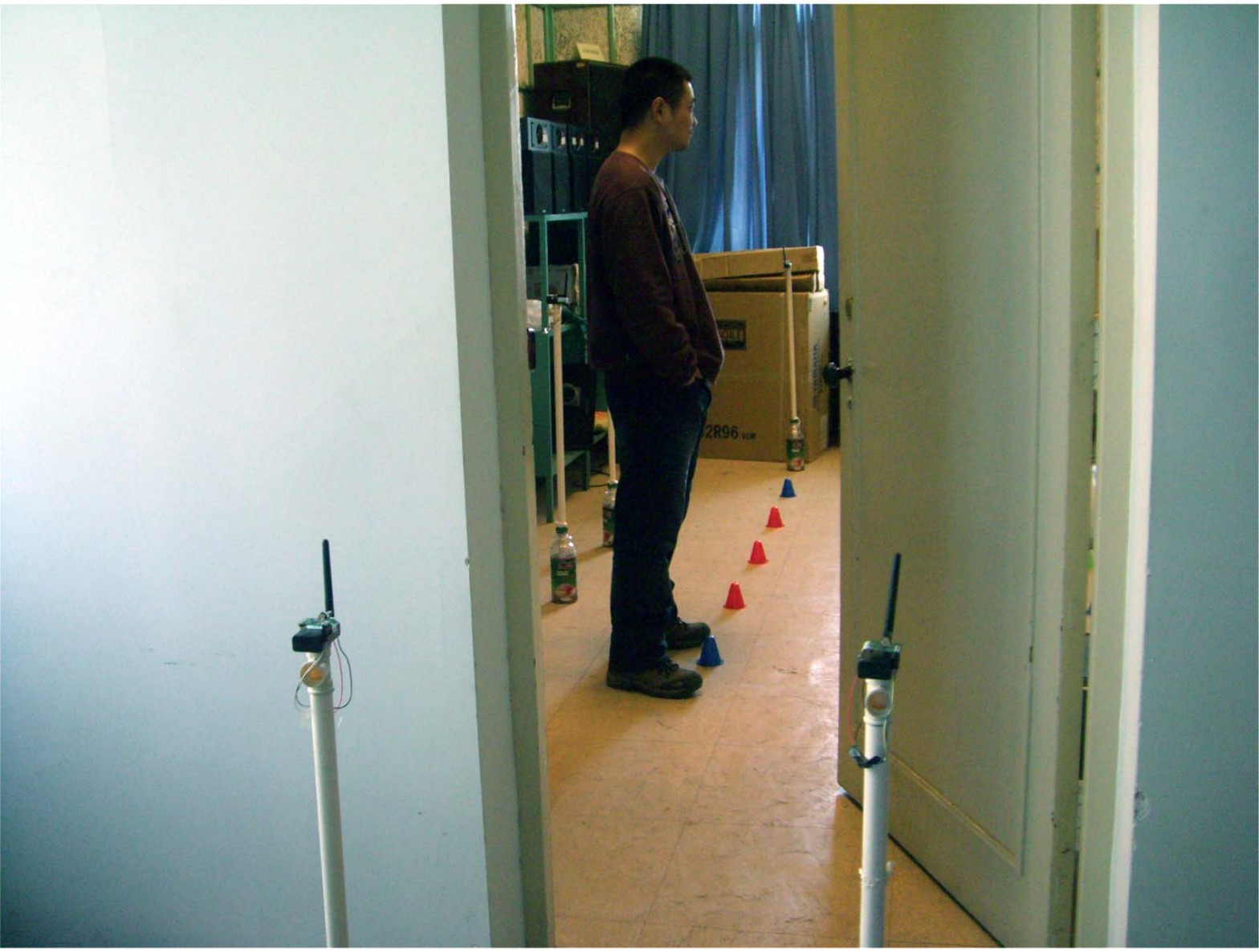}}
\subfigure[Node positions.]{
\includegraphics[height = 3.5cm]{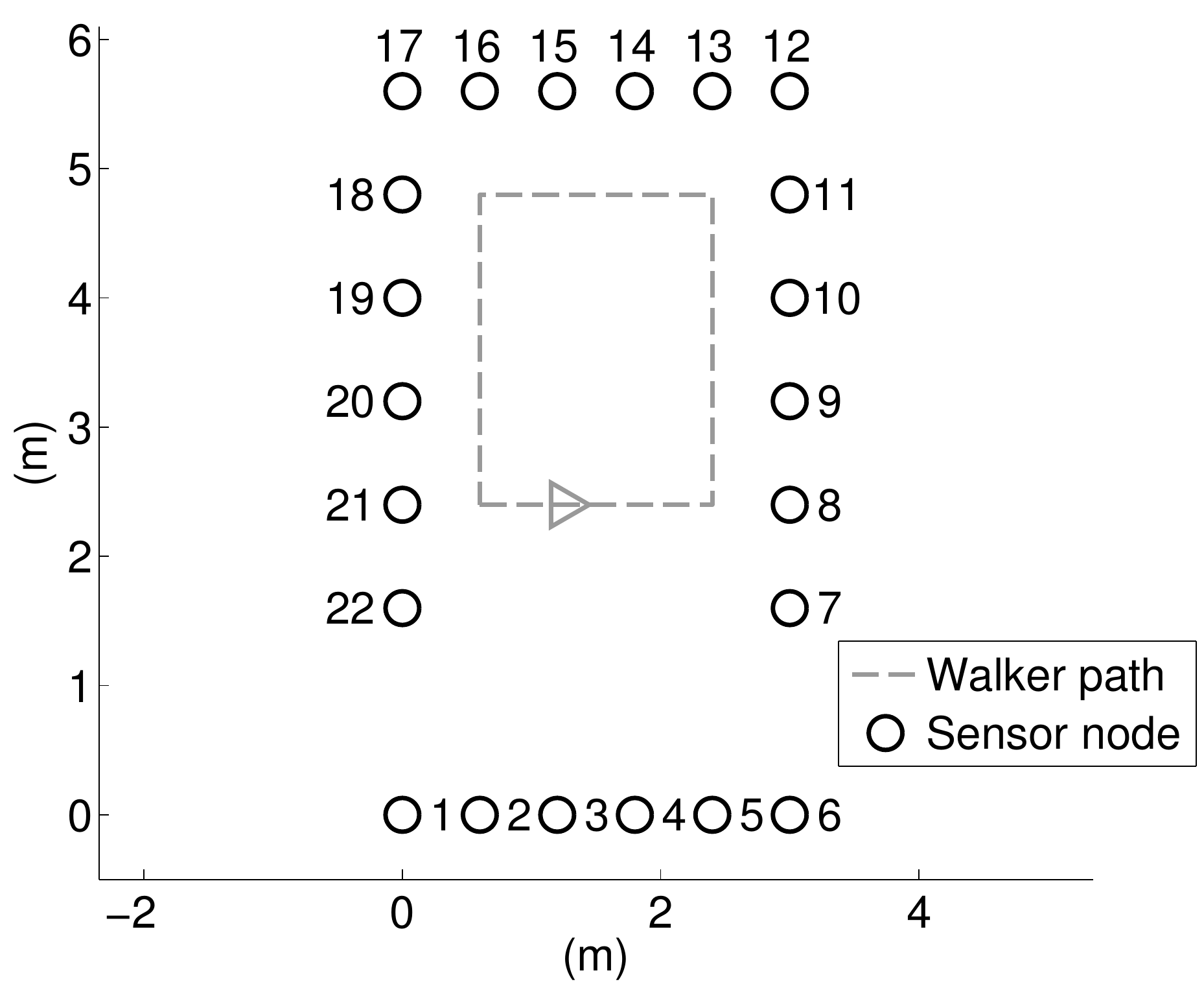}
\label{fig:Experiment3SetupDiagram}}
\caption[Node deployment for the through-wall experiment]{Node deployment for experiment set 3.}
\label{fig:Experiment3Setup}
\end{figure}

The third set of experiments was also conducted in an indoor lab.  This time, some of the nodes were placed in a hallway outside the lab (see Fig.~\ref{fig:Experiment3Setup}), thus monitoring the inside of the lab through a solid wall which passed through the gap between nodes~6 and~7 and~1 and~22 in Fig.~\ref{fig:Experiment3SetupDiagram}.  This scenario is one of the hardest for RF sensing applications to cope with.

The data from these experiments allow us to test algorithms in several different environments.

\subsection{Performance Evaluation Methods}\label{sec:PerformanceEvaluationMethods}

One way to evaluate our algorithms would be to compare the links they assign to the foreground set $\mathcal{F}$ with ground truth knowledge of which links are actually in the foreground.  However, there is no way to know this.  One might assume that links whose LOS paths intersect the obstruction ought to be in the foreground, but this is not guaranteed, especially indoors where multipath effects are non-trivial and the non-line-of-sight (NLOS) components of a signal can be heavy contributors to its RSS value.  If one of these NLOS components is obstructed, the associated link may well be in the foreground. {\color{\changedText}We therefore evaluate our algorithms in two ways.  We first look at their ability to estimate the baseline RSS measured during the offline calibration period.  Then, to see how useful our values are to a real-world application, we examinine the performance of an RF tomographic tracking algorithm initialized with our values.}

\subsubsection[Comparison With the Measured Baseline RSS]{Metric 1: Comparison With Measured $\mathbf{R_B}$}\label{sec:EvaluateRMSE}

Because we have access to direct measurements of the baseline RSS from the calibration period of each experiment set, we define $\tilde{R}_B[\ell]$ as the mean RSS measured on $\ell$ over this time period, when the region of interest was intentionally left vacant.  Recall that this is the data which existing RF tomographic algorithms already assume is available to them.

Our estimate of the baseline RSS on link $\ell$, $\hat{R}_B[\ell]$, is then calculated according to
\begin{equation}
\label{eq:RHat}
\hat{R}_B[\ell] = \frac{\sum_{k\in K}\psi^{(k)}[\ell]R^{(k)}[\ell]}{\sum_{k\in K}\psi^{(k)}[\ell]}
\end{equation}
where $K$ is the total number of frames available at a given time and $\psi^{(k)}[\ell]$ is an indicator function
\setlength{\arraycolsep}{0.0em}
\begin{equation}
\label{eq:Indicator}
\psi^{(k)}[\ell] = \begin{cases}
1, & \text{if }\ell\in\mathcal{B}\text{ in frame }k \\
0, & \text{if }\ell\in\mathcal{F}\text{ in frame }k
\end{cases},
\end{equation}
\setlength{\arraycolsep}{5pt}built using the output of our background subtraction algorithms.  This $\psi^{(k)}[\ell]$ will take different values depending on whether background subtraction is carried out using TBM, FABS or FABS-MMCL.

Once we obtain $\hat{R}_B[\ell]$, we can calculate the root mean square estimation error of the approximation (measured in dBm/link) such that
\begin{equation}
\label{eq:EstimationError}
\text{Estimation Error} = \sqrt{\frac{1}{\Lambda}\sum_{\ell\in\Lambda}\left(\tilde{R}_B[\ell]-\hat{R}_B[\ell]\right)^2}
\end{equation}
where $\Lambda$ is the total number of links in the network. This provides us with a numerical qualification for how well we have estimated the baseline RSS, which can easily be compared across our different algorithms.  We can also calculate this metric for the mean approximation (MA) method described in Section~\ref{chapter:Ch4}, which constitutes the simplest possible way to estimate baseline RSS without access to a calibration period.  Note that the MA algorithm corresponds to setting $\psi^{(k)}[\ell]=1$ for all values of $k$ and $\ell$.

\subsubsection[Tracking Performance]{Metric 2: Tracking Performance}\label{sec:EvaluateTracking}

{\color{\changedText}To get an intuitive feel for the implications of a certain level of estimation error, we use our data in a real-world RF sensing network application.  Many such applications exist, but we choose to employ Li et al.'s RF tomographic tracking algorithm~\cite{Li2011}, which must be given a value for the baseline RSS on each link in the network before it can begin tracking.}  By comparing tracking performance when using $\mathbf{\tilde{R}_B}$, the vector of measured baseline RSS values, to tracking performance when using different versions of $\mathbf{\hat{R}_B}$, we can see whether our estimates of the baseline RSS are accurate enough for real-world applications.

Because Li et al.'s algorithm is non-deterministic, for each value of $\mathbf{\tilde{R}_B}$ or $\mathbf{\hat{R}_B}$, we run $A = 100$ realizations of the algorithm and report the performance over this ensemble in the form of the root mean square tracking error calculated as
\begin{equation}
\label{eq:TrackingError}
\text{Tracking Error} = \frac{1}{A}\sum_{a=1}^{A}{\sqrt{\frac{1}{T}\sum_{t=1}^{T}{\left|\left|\mathbf{z}_t-\mathbf{\hat{z}}^{(a)}_t\right|\right|^2}}}
\end{equation}
where $\mathbf{\hat{z}}^{(a)}_t$ is the estimated value of the person's position at time $t\in T$ as reported by realization $a$.

\subsubsection{Additional Evaluation}\label{sec:AdditionalEvaluation}

To further understand the behaviour of our background subtraction algorithms, we look at which links and RSS values are selected for the foreground using histograms and other diagrams. {\color{\changedText}We also investigate the algorithms' run-times in Section~\ref{sec:Performance}.}

Because this work constitutes the first examination of the potential of background subtraction to estimate baseline RSS in WSNs, we are mostly interested in determining the maximum improvements which can be gained from this technique.  {\color{\changedText}Therefore, we present results which are obtained when the background subtraction parameters ($\theta$, $N$, etc.) are specifically those found (via a series of parameter sweeps) to give the best performance and lowest estimation error}.  To offset this, in Section~\ref{sec:Parameters}, we provide an examination of the algorithms' sensitivity to parameter values which differ from these settings.

\subsection{Results for Each Experiment Sets}\label{sec:Results}
\subsubsection[Results for the Experiment Set 1]{Results for Experiment Set~1}\label{sec:Results1}

\begin{figure}
\centering
\includegraphics[width=3in]{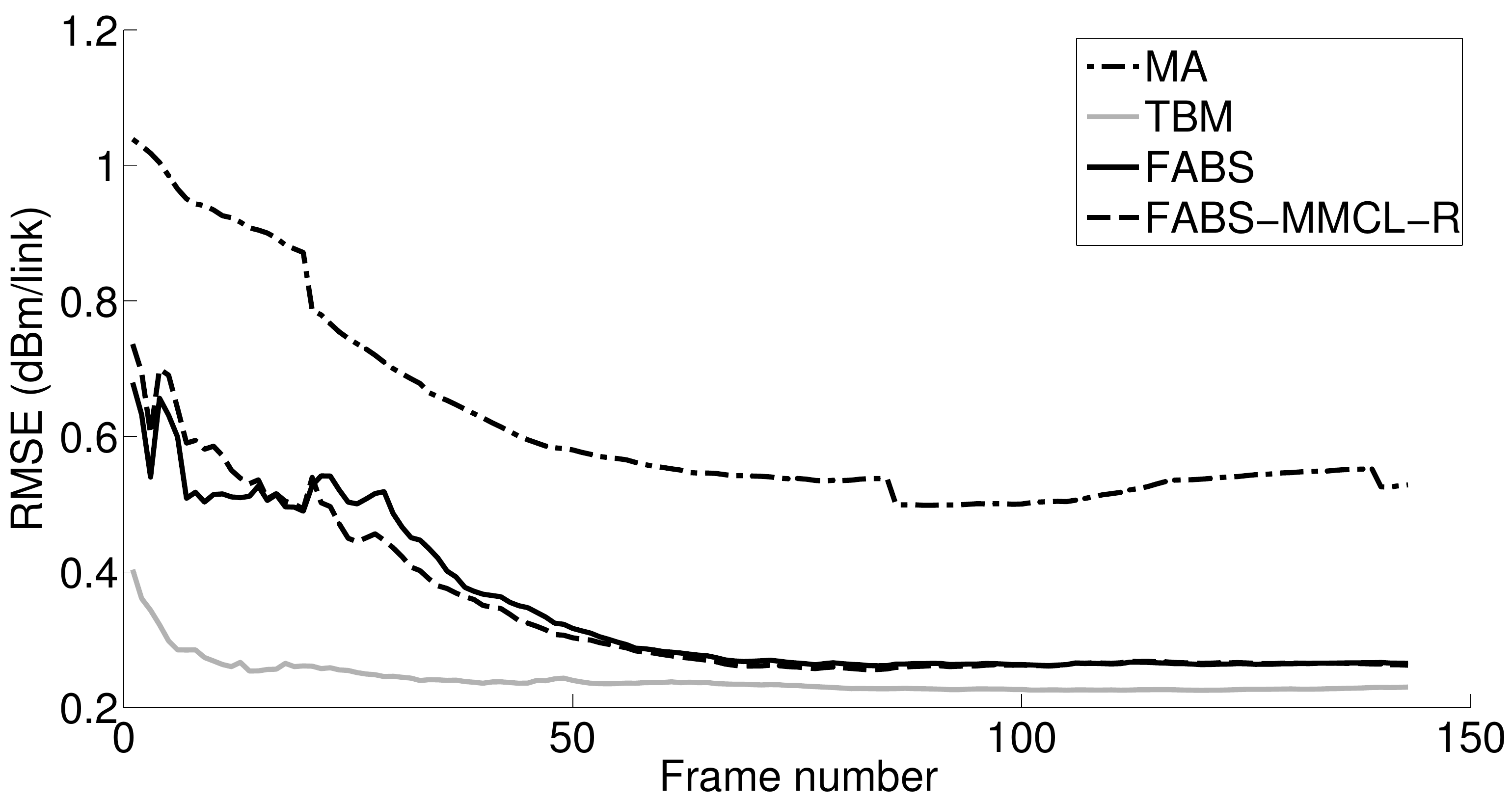}
\caption[Estimation error for experiment set 1]{RMSE obtained when estimating the baseline RSS for experiment set~1.  The~143 frames used were collected while a walker made one round trip around the set path over a period of~$\approx$~20 seconds.}
\label{fig:Experiment1RMSE}
\end{figure}

Fig.~\ref{fig:Experiment1RMSE} presents the estimation error obtained when using our background subtraction algorithms to estimate baseline RSS values for an experiment conducted in an outdoor field.  Table~\ref{tab:Experiment1Parameters} presents the parameter values used to obtain the pictured results. 

\begin{table}
\renewcommand{\arraystretch}{1.3}
\caption[Parameter values used for experiment set 1]{Parameter values used for experiment set~1.}
\label{tab:Experiment1Parameters}
\centering
\begin{tabular}{c||c||c||c}
	\hline
	Parameter              & TBM  & FABS & FABS-  \\
	                       &       &      & MMCL-R \\
  \hline\hline                                         
  $N$                    & 25   & 25    & 25     \\
  \hline
  $\sigma_t^2$           & 4    & 4     & 4      \\
  \hline
  $\theta$               & 0.17 & 0.17  & 0.17   \\
  \hline
  $\tau$                 & --   & 0     & 0      \\
  \hline
  $\sigma_s^2$           & --   & 4     & 4      \\
  \hline
  $\eta$                 & --   & 1     & 1      \\
  \hline
  $\gamma$               & --   & --    & 5      \\
  \hline
  $C$                    & --   & --    & 95     \\
  \hline
\end{tabular}
\end{table}

{\color{\changedText}From Fig.~\ref{fig:Experiment1RMSE}, we see that MA is fairly accurate already and that TBM provides another sizable improvement compared to MA.  Recall that these are the least computationally-complex of all the algorithms we present and that no knowledge of the nodes' positions is required to apply them.  Consequently, it is particularly encouraging that they are this successful.}  We also see that the more complicated algorithms do not actually do any better than TBM in this case.

Since this original experiment set includes only a single ``walk'', we validate these results by repeating our analysis on data sets collected several months later, in a different outdoor field also located at BUPT; these results are shown in~\cite{Edelstein2011}. Note that in both fields, we see the same pattern in our results: TBM shows an improvement over MA while the more complex FABS-based approaches do not show any significant improvement compared to TBM.  Additionally, it is worth noting that these similar results are obtained in two different outdoor environments with nodes set out in two slightly different patterns  (since the broken nodes~11 and~13 from experiment set~1 were fixed before more experiments were conducted), but using the same parameters seen in Table~\ref{tab:Experiment1Parameters}.  This is very encouraging, because even if these parameter values seem slightly arbitrary, this shows that at least they are transferrable between similar environments.

Returning to the experiment set~1 data, we also use Li et al.'s tracking algorithm~\cite{Li2011} (initialized in turn with $\mathbf{\tilde{R}_B}$ and with various values of $\mathbf{\hat{R}_B}$) to track a walking person.  Tracking errors are shown in Table~\ref{tab:Experiment1RMSE}.

\begin{table}
\renewcommand{\arraystretch}{1.3}
\caption[Tracking error for experiment set 1]{RMSE for tracking using experiment set~1.}
\label{tab:Experiment1RMSE}
\centering
\begin{tabular}{c||c}
	\hline
	 Algorithm & Tracking Error (m)\\
	 \hline\hline
	 Calibration Data & 0.4073$\pm$0.0655 \\
	 \hline
	 MA & 0.4363$\pm$0.0711 \\
	 \hline
	 TBM & 0.4319$\pm$0.0655 \\
	 \hline
	 FABS & 0.4335$\pm$0.0667 \\
	 \hline
	 FABS-MMCL-R & 0.4339$\pm$0.0616 \\
  \hline
\end{tabular}
\end{table}

\begin{figure}
\centering
\subfigure[TBM foreground]{
\includegraphics[width=2.61cm]{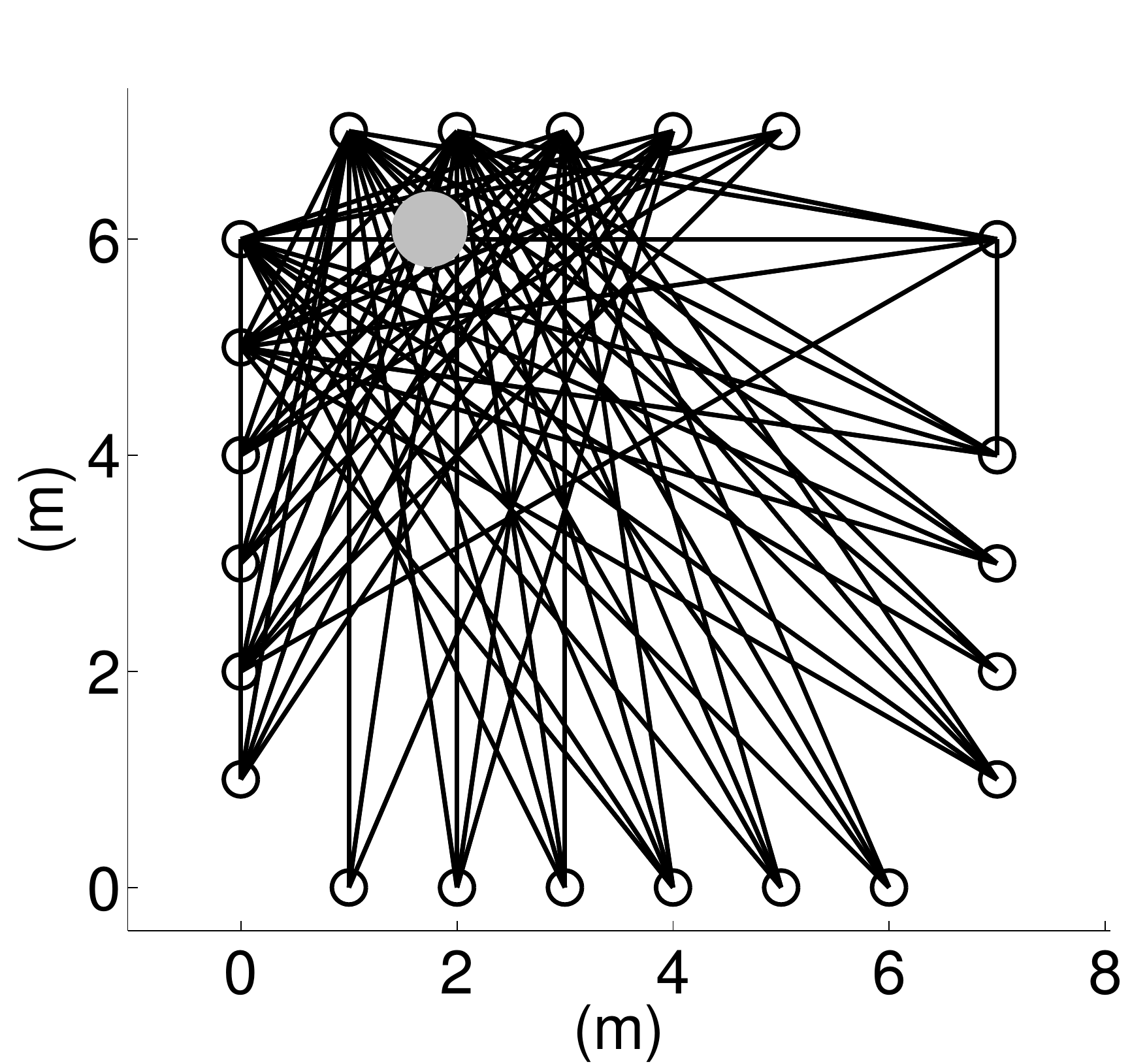}}
\subfigure[FABS foreground]{
\includegraphics[width=2.61cm]{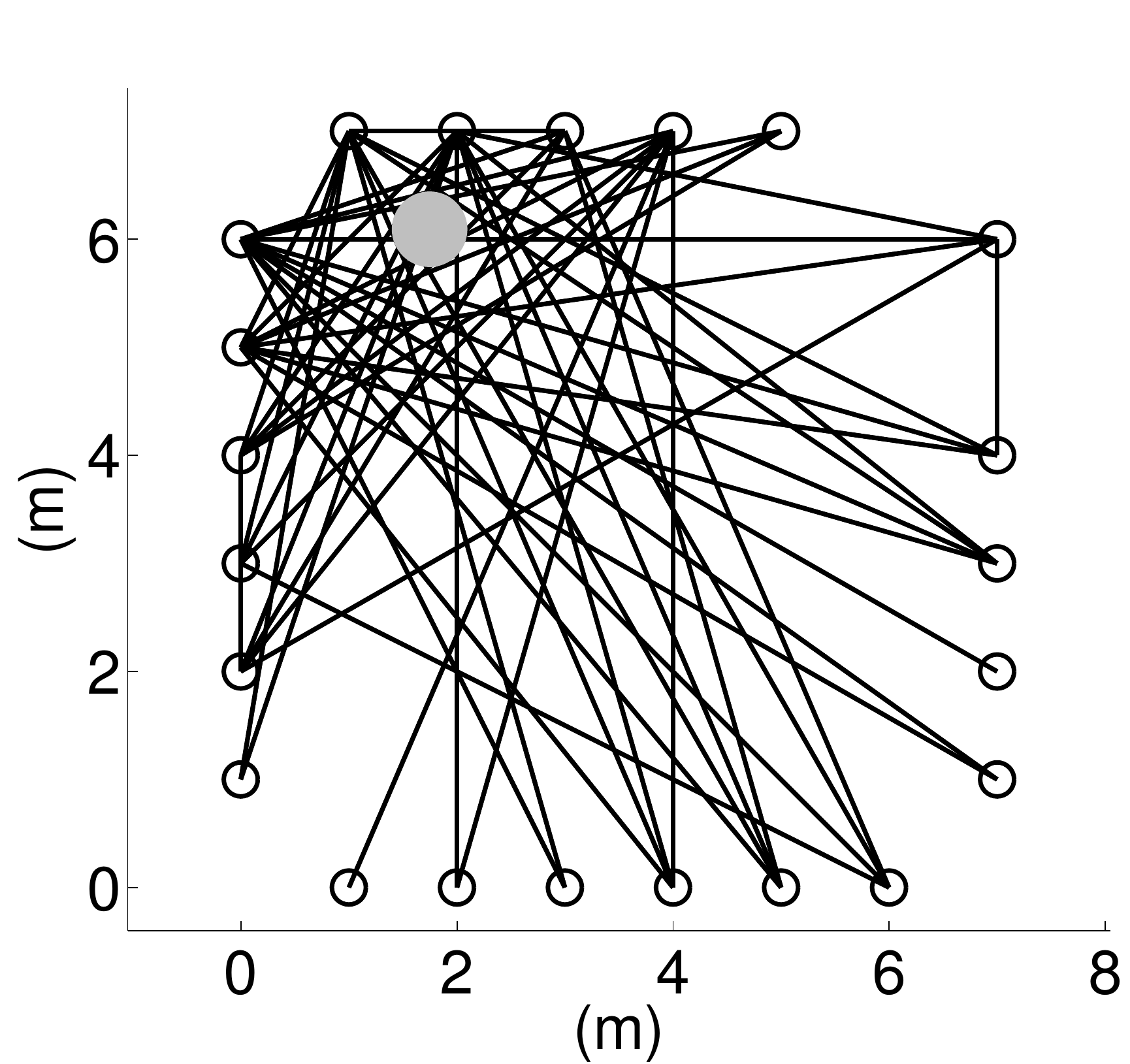}}
\subfigure[FABS-MMCL-R foreground]{
\includegraphics[width=2.61cm]{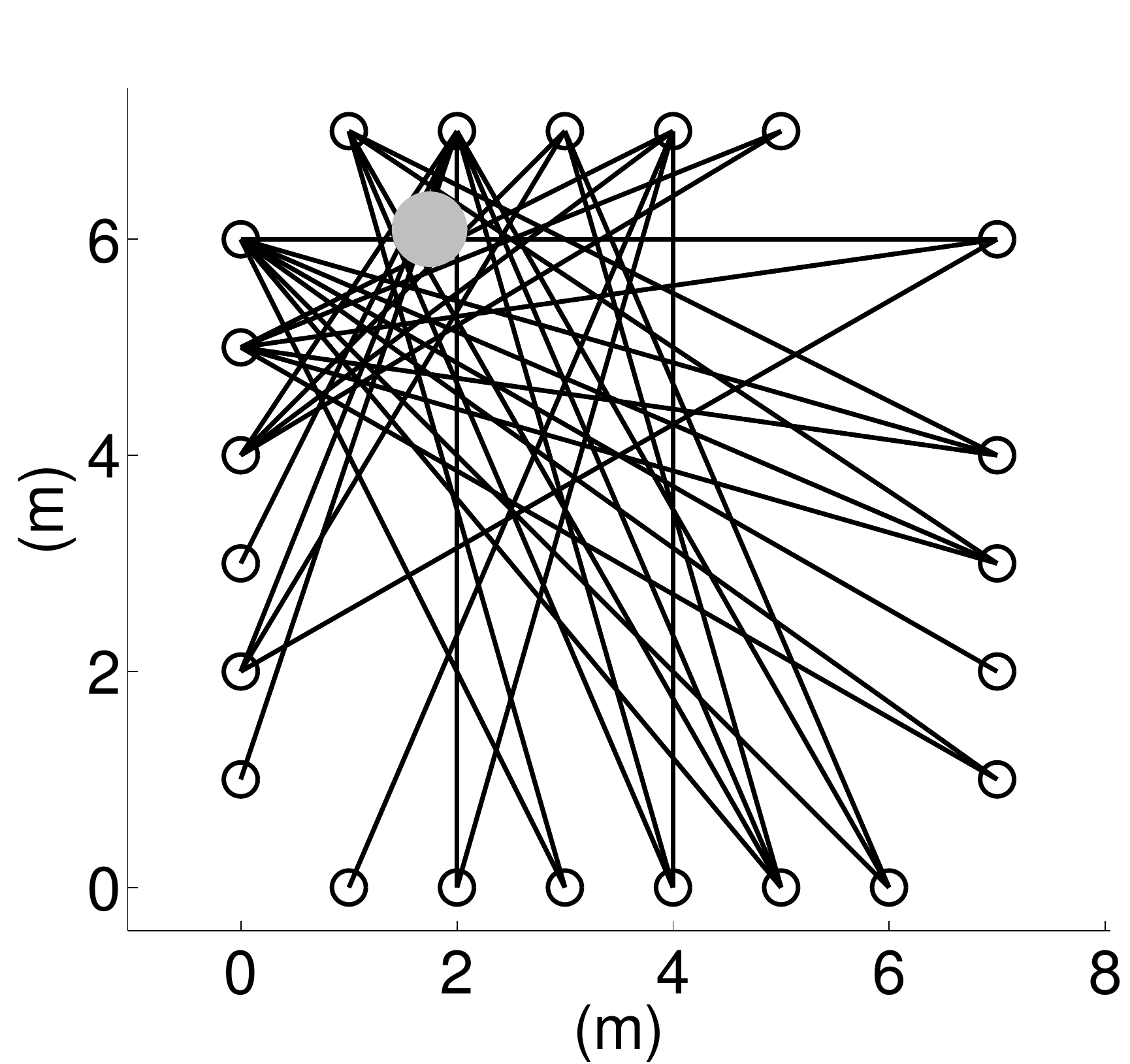}} \\
\subfigure[TBM background]{
\includegraphics[width=2.61cm]{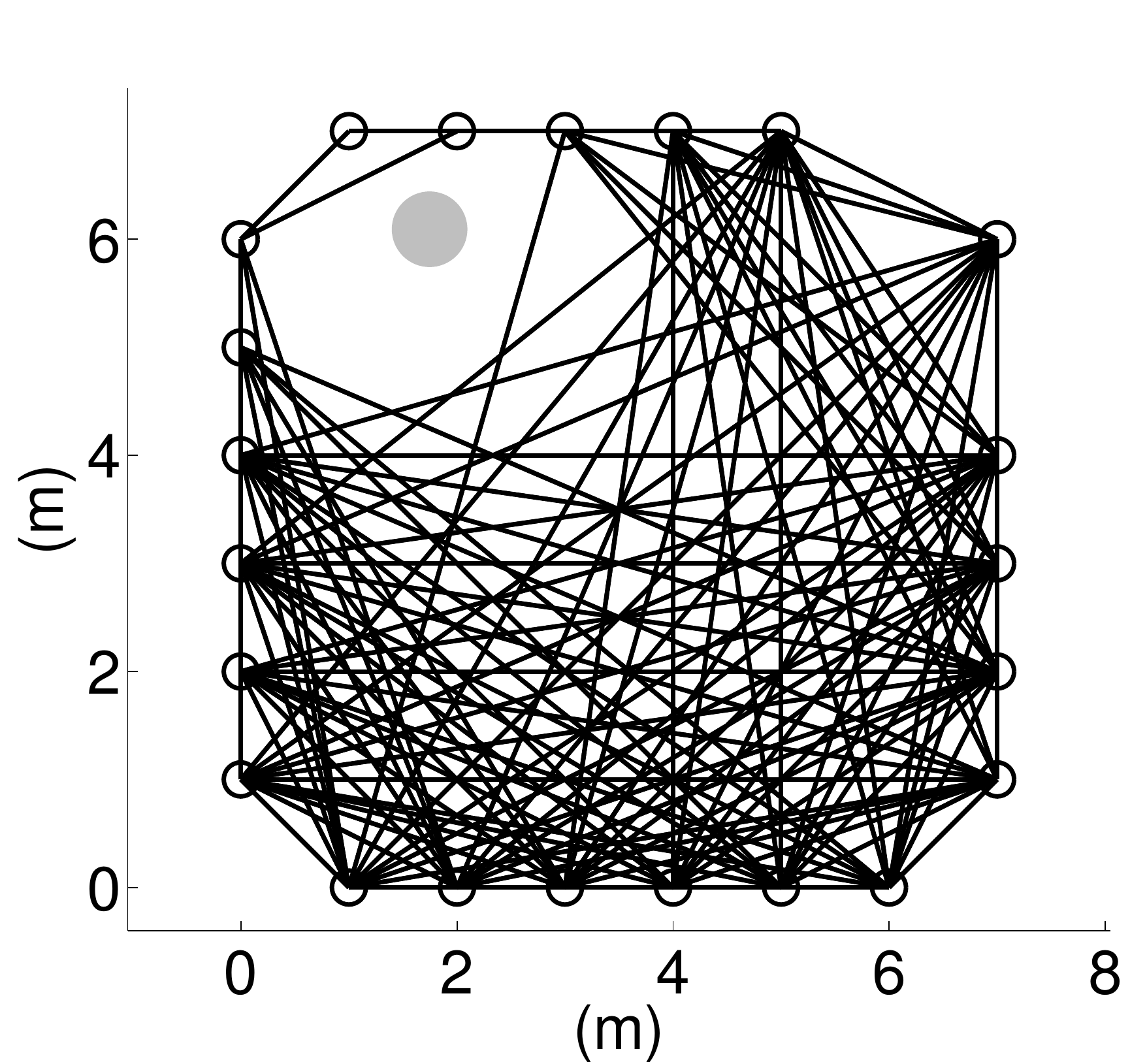}}
\subfigure[FABS background]{
\includegraphics[width=2.61cm]{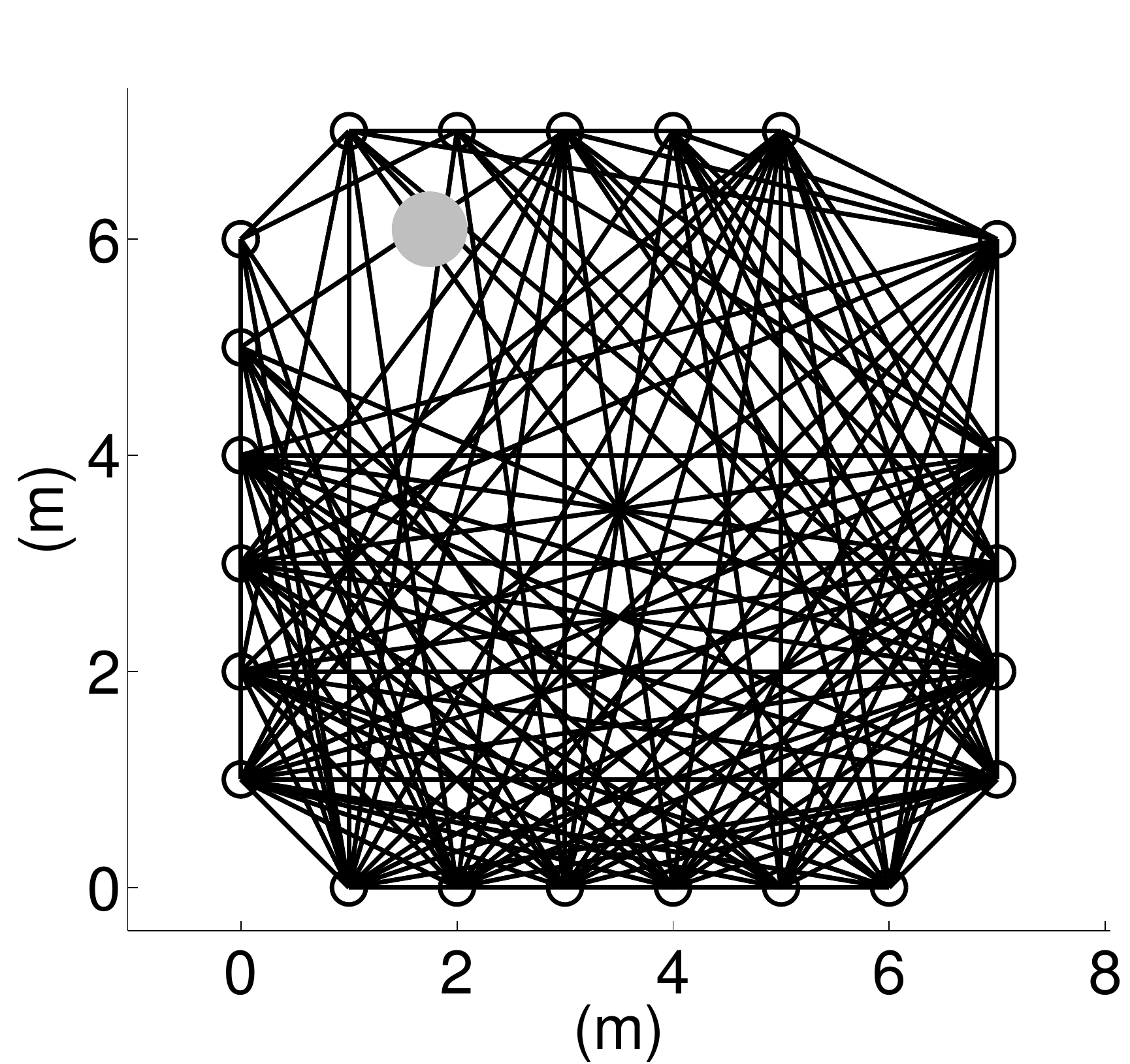}}
\subfigure[FABS-MMCL-R background]{
\includegraphics[width=2.61cm]{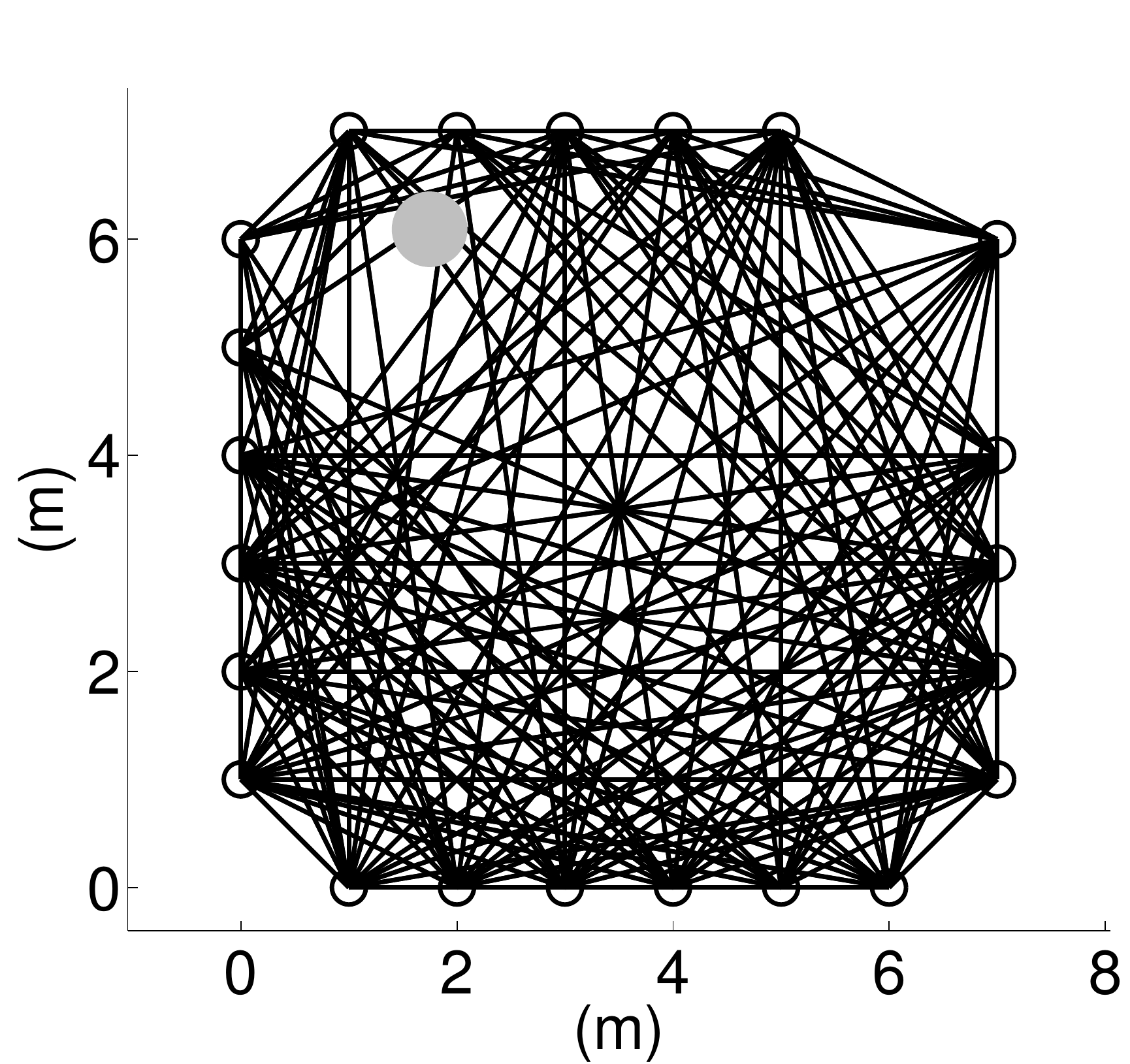}}
\caption{Links assigned to the foreground and background by various background subtraction algorithms for a single {\color{\changedText}representative} frame in experiment set~1.  The known position of the obstruction during this frame is marked.}
\label{fig:PlayAgainExperiment1}
\end{figure}

From these results, we can see that the ideal situation (where we initialize tracking with measured data) provides the best performance, as expected.  {\color{\changedText}We can also see that using MA to build $\mathbf{\hat{R}_B}$ causes our performance to suffer only slightly, while TBM provides a small improvement over MA---though even this small improvement may well be desirable in systems intended to replace video or RFID tracking systems.  Finally, we see that the more complex algorithms are no more helpful than TBM. In all cases, the tracking error is low, lending support to the idea that any of the four methods shown can serve as an excellent substitute to measuring $\mathbf{R_B}$ directly.  This makes sense in an outdoor environment; the situation will change in more complex, indoor environments.

To understand why FABS and FABS-MMCL do not improve over MA and TBM, we examine which links are being selected as foreground and background links by each algorithm. This is seen (for a single representative frame) in Fig.~\ref{fig:PlayAgainExperiment1}.}  Note that any link which is not in the foreground in a given frame must be in the background, so the bottom row of figures is the logical negation of the top row.

As we described in Section~\ref{sec:FABSMMCL}, FABS is designed to grow the set of foreground links created by TBM.  However, in Fig.~\ref{fig:PlayAgainExperiment1}, we see that, in this case, FABS is \emph{removing} links from the set created by TBM.  We might hypothesize that this is occurring because TBM is being overly-greedy, perhaps due to $\theta$ being set too high.  Recall though, that the parameters have all been adjusted to optimize performance.  Looking at Fig.~\ref{fig:PlayAgainExperiment1} just provides more supporting evidence, then, that for outdoor environments, the low-complexity TBM algorithm is already doing an admirable job identifying foreground links, and any attempt to improve on its efforts using FABS or FABS-MMCL is futile.

\begin{figure}
\centering
\subfigure[All RSS measurements]{
\includegraphics[width=2.5in]{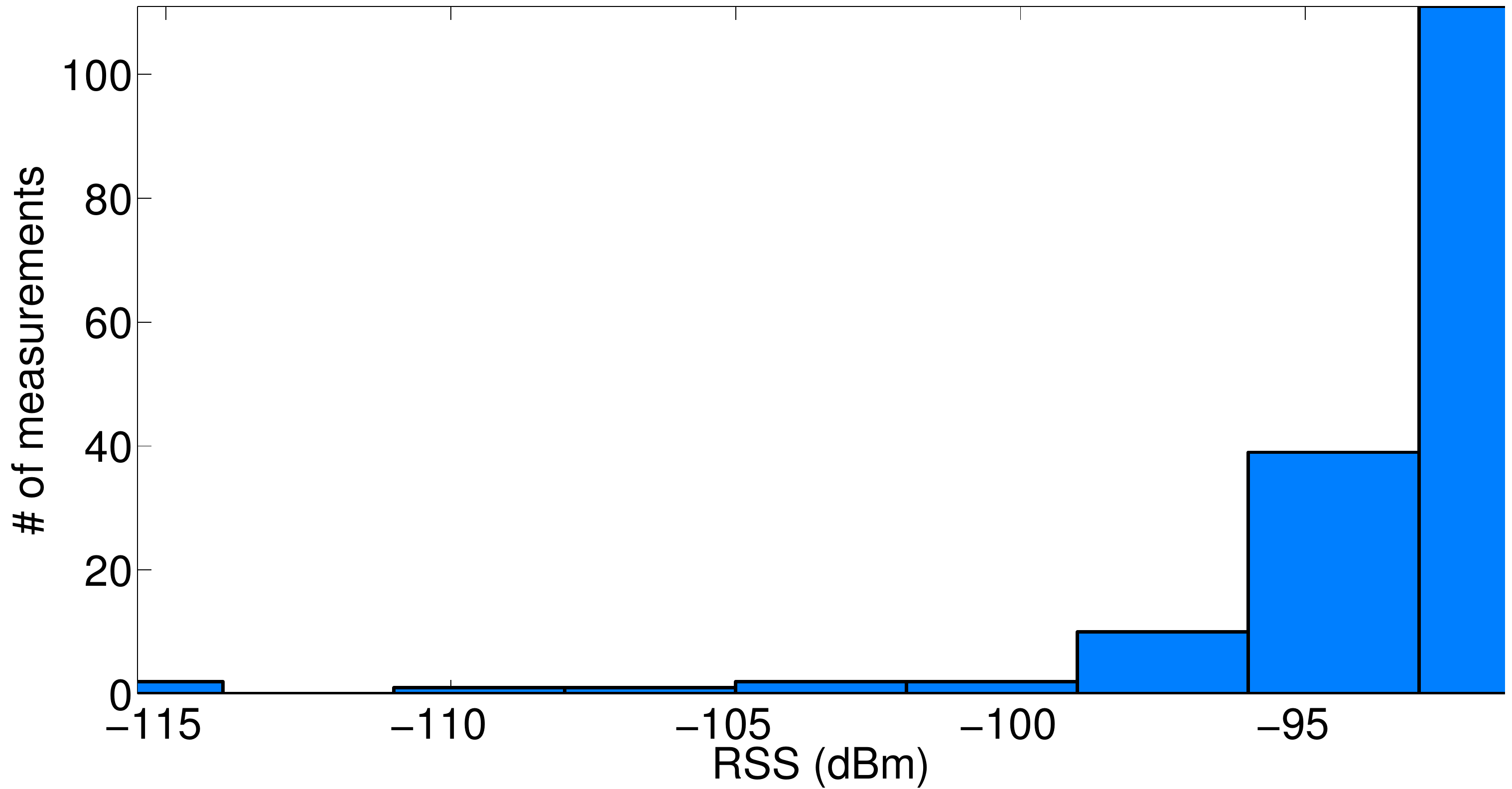}
\label{fig:HistogramExperiment1All}}
\subfigure[Background RSS measurements as determined by TBM]{
\includegraphics[width=2.5in]{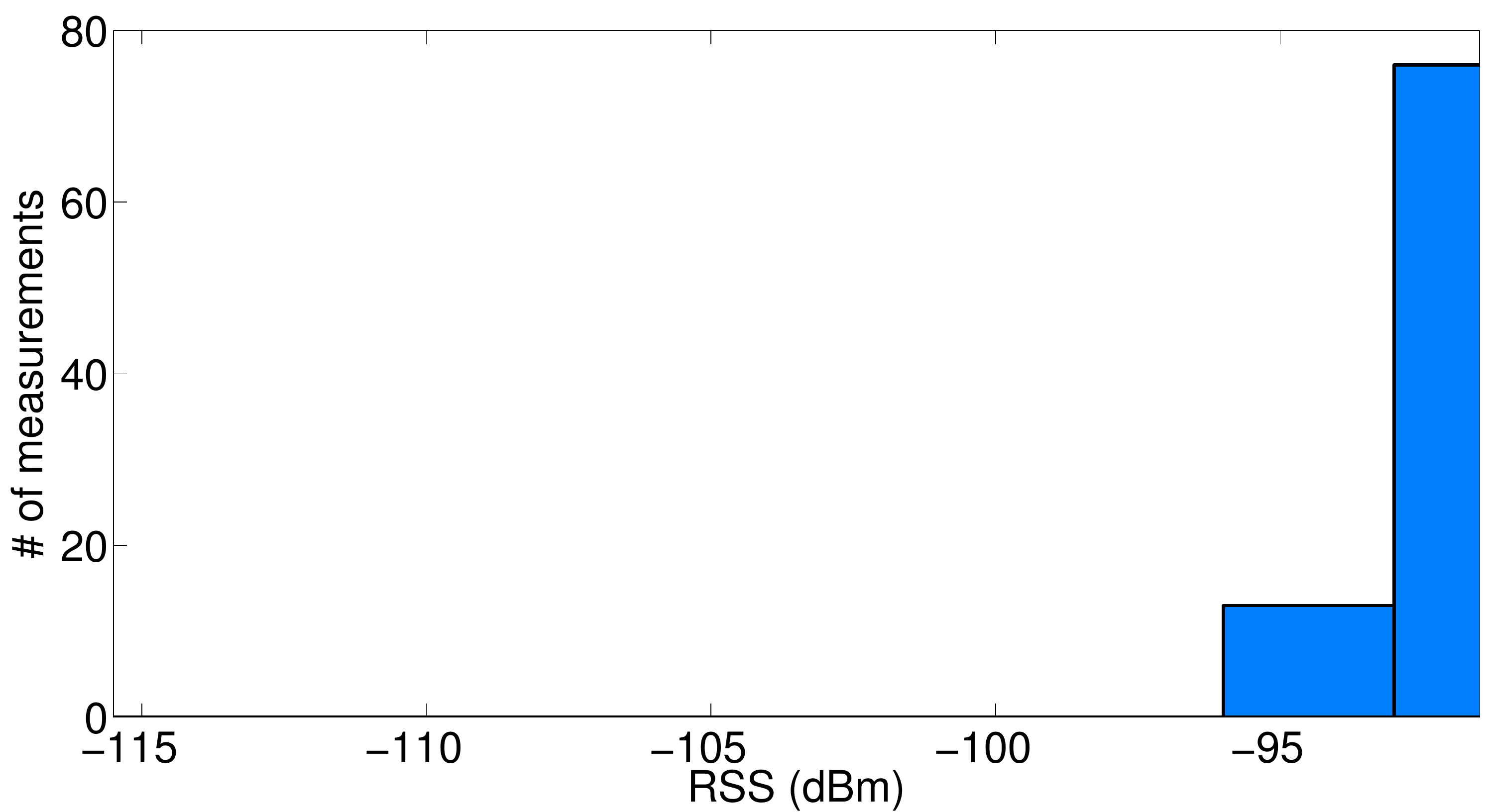}
\label{fig:HistogramExperiment1BG}}
\caption[Histogram of RSS measurements for a single link]{Histogram of RSS measurements seen on the link connecting nodes at $(7,1)$ and $(1,0)$ during experiment set~1.}
\label{fig:HistogramExperiment1}
\end{figure}

Having examined the background vs. foreground behaviour of all the links in the network for a single frame, we will also look at the behaviour of one link in the network over all frames.  In Fig.~\ref{fig:HistogramExperiment1All}, we see a histogram of the RSS measured on a particular link in the WSN (in this case, the link connecting nodes at $(7,1)$ and $(1,0)$) over all frames in $K$, and in Fig.~\ref{fig:HistogramExperiment1BG}, we see a histogram of the subset of these values deemed to be in the background by TBM.  From these figures, we see that the outliers from Fig.~\ref{fig:HistogramExperiment1All} are properly identified as being in the foreground and are removed in Fig.~\ref{fig:HistogramExperiment1BG}.   
This behaviour is representative of that seen on all links in the network.

\subsubsection[Results for Experiment Set 2]{Results for Experiment Set~2}\label{sec:Results2}

\begin{figure}
\centering
\includegraphics[width=3in]{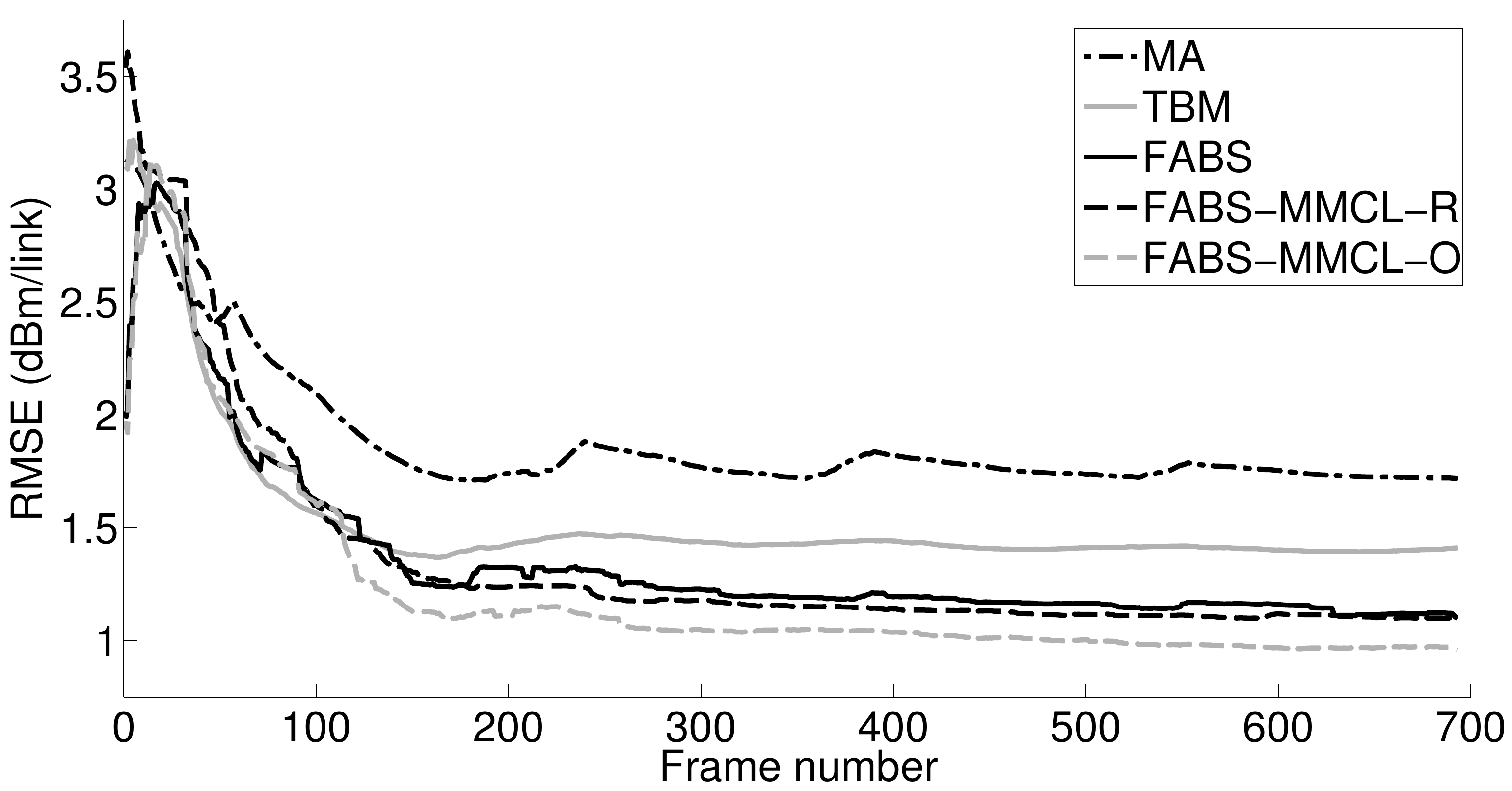}
\caption[Estimation error for experiment set 2]{RMSE obtained when estimating the baseline RSS for experiment set~2.  The~693 frames used were collected while a walker made four round trips around the set path over a period of~$\approx$~80 seconds.}
\label{fig:Experiment2RMSE}
\end{figure}

Fig.~\ref{fig:Experiment2RMSE} presents the root mean square estimation error obtained when using our background subtraction algorithms to estimate baseline RSS values for an experiment conducted in an indoor lab.  Table~\ref{tab:Experiment2Parameters} presents the associated parameter values.

\begin{table}
\renewcommand{\arraystretch}{1.3}
\caption[Parameter values used for experiment set 2]{Parameter values used for experiment set~2.}
\label{tab:Experiment2Parameters}
\centering
\begin{tabular}{c||c||c||c||c}
	\hline
	Parameter              & TBM  & FABS & FABS-  & FABS-  \\
	                       &       &      & MMCL-R & MMCL-O \\
  \hline\hline                   
  $N$                    & 35   & 35    & 35     & 35     \\
  \hline
  $\sigma_t^2$           & 17   & 17    & 17     & 17     \\
  \hline
  $\theta$               & 0.05 & 0.05  & 0.05   & 0.05   \\
  \hline
  $\tau$                 & --   & 0.75  & 0.75   & 0.75   \\
  \hline
  $\sigma_s^2$           & --   & 10    & 10     & 10     \\
  \hline
  $\eta$                 & --   & 5     & 5      & 1      \\
  \hline
  $\gamma$               & --   & --    & 25     & 50     \\
  \hline
  $C$                    & --   & --    & 35     & --     \\
  \hline
\end{tabular}
\end{table}

If we compare Fig.~\ref{fig:Experiment2RMSE} to the similar figure generated for experiment set 1, we can see some important differences.  First, if we compare the performance of MA in each case, we see that even this simple approach performs much better in the outdoor environment.  This hints that our task of properly estimating the baseline RSS is far harder indoors than it is outdoors.  Fortunately, TBM provides an improvement over MA in the indoor case and, moreover, FABS and FABS-MMCL provide further improvement still (with FABS-MMCL-O providing more improvement than FABS-MMCL-R).  Even after applying these algorithms, the lowest estimation error we achieve is still higher than the estimation error seen for the outdoor experiment sets, but this is understandable given the increased complexity and more intense multipath effects inherent in an indoor environment; additional results showing similar behavior are presented in~\cite{Edelstein2011}.

We can see another result of these multipath effects if we compare the convergence rates between the indoor and outdoor experiment sets.  This convergence rate depends on the amount of noise affecting the RSS measurements and, in particular, on how often we have access to background measurements for each link.  At one extreme, if there is absolutely no measurement noise and if no links are ever deemed to be in the foreground, $\mathbf{\hat{R}_B}$ can converge after a single frame.  On the other hand, if many links are in the foreground in each frame (e.g., because there are many obstructions, because the obstruction(s) are very large or because multipath effects cause the obstruction(s) to affect many links) or if the same links are in the foreground for many frames (again, because the obstruction(s) affect many links or because they do not move around very much), it may take several frames before we can obtain a background measurement for each link.  The estimate $\mathbf{\hat{R}_B}$ will then converge more slowly.

In our case, the same size obstruction was moving at approximately the same rate in both the indoor and outdoor sets of experiments.  However, due to multipath effects, more links are in the foreground for each frame in the indoor environment (as will be seen when we graphically examine the foreground links selected for a single frame).  Therefore, fewer measurements are available for use in calculating $\mathbf{\hat{R}_B}$ at any given time, causing slower convergence.

Again, we also use Li et al.'s tracking algorithm~\cite{Li2011} to detect the movements of a person walking around the lab. The tracking errors seen over~100 realizations of the algorithm (initialized with $\mathbf{\tilde{R}_B}$ and $\mathbf{\hat{R}_B}$) are shown in Table~\ref{tab:Experiment2RMSE}.

\begin{table}
\renewcommand{\arraystretch}{1.3}
\caption[Tracking error for experiment set 2]{RMSE for tracking using experiment set~2.}
\label{tab:Experiment2RMSE}
\centering
\begin{tabular}{c||c}
	\hline
	 Algorithm & Tracking Error (m) \\
	 \hline\hline
	 Calibration Data & 0.6060$\pm$0.0022 \\
	 \hline
	 MA & 0.6801$\pm$0.1763 \\
	 \hline
	 TBM & 0.6404$\pm$0.1962 \\
	 \hline
	 FABS & 0.5994$\pm$0.0033 \\
	 \hline
	 FABS-MMCL-R & 0.6221$\pm$0.0841 \\
  \hline
  FABS-MMCL-O & 0.6120$\pm$0.0956 \\
  \hline
\end{tabular}
\end{table}

As expected, using MA in place of the measured $\mathbf{\tilde{R}_B}$ increases the tracking error from 0.6060~m to 0.6801~m.  Simple TBM reduces this error to 0.6404~m while the more complicated FABS-based approaches all reduce the error closer to the level of (or below the level of!) the error seen when $\mathbf{\tilde{R}_B}$ is used. {\color{\changedText}The error variance is also reduced using more complex methods.}

\begin{figure}
\centering
\subfigure[TBM foreground]{
\includegraphics[width=2.57cm]{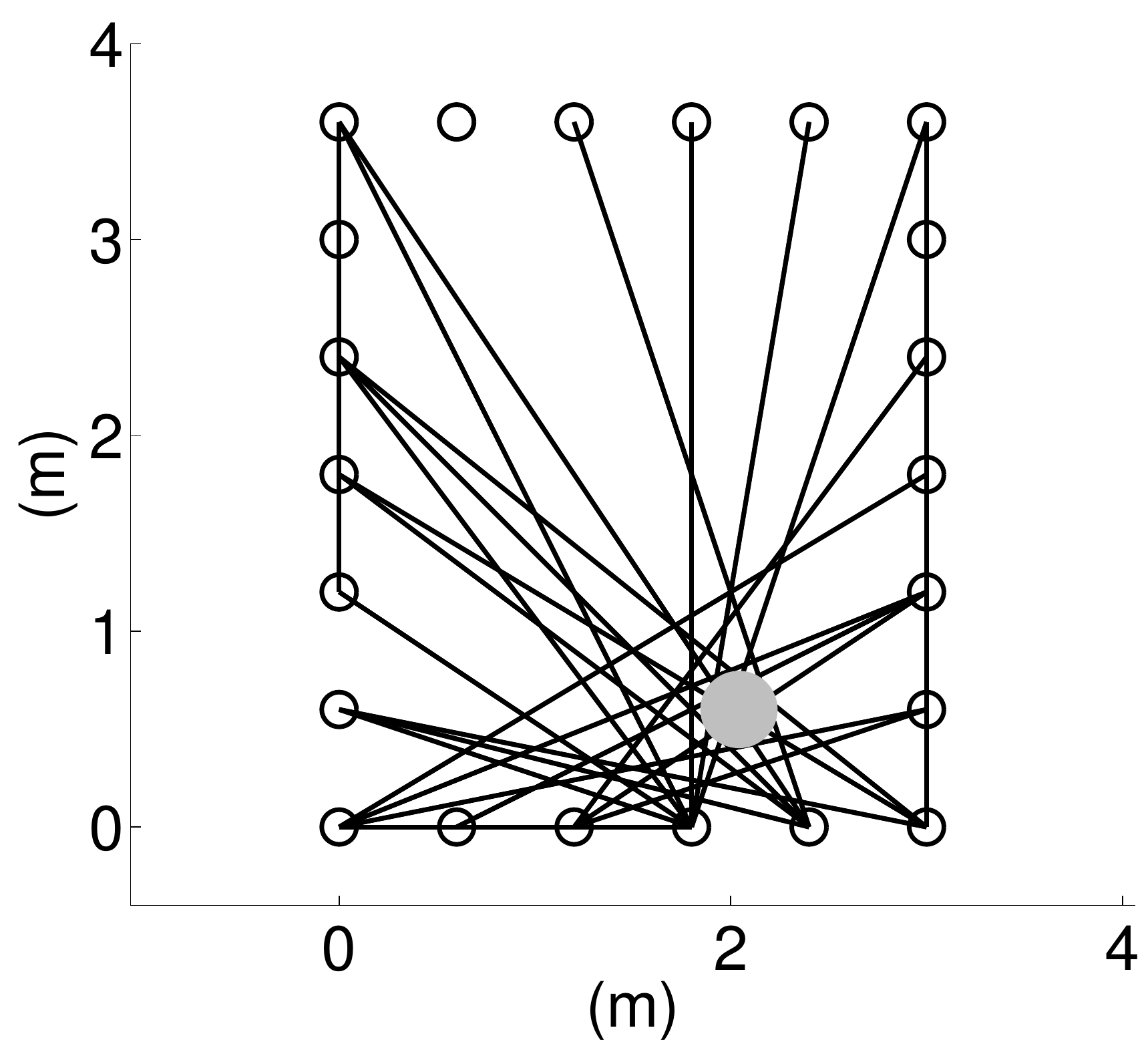}}
\subfigure[FABS foreground]{
\includegraphics[width=2.57cm]{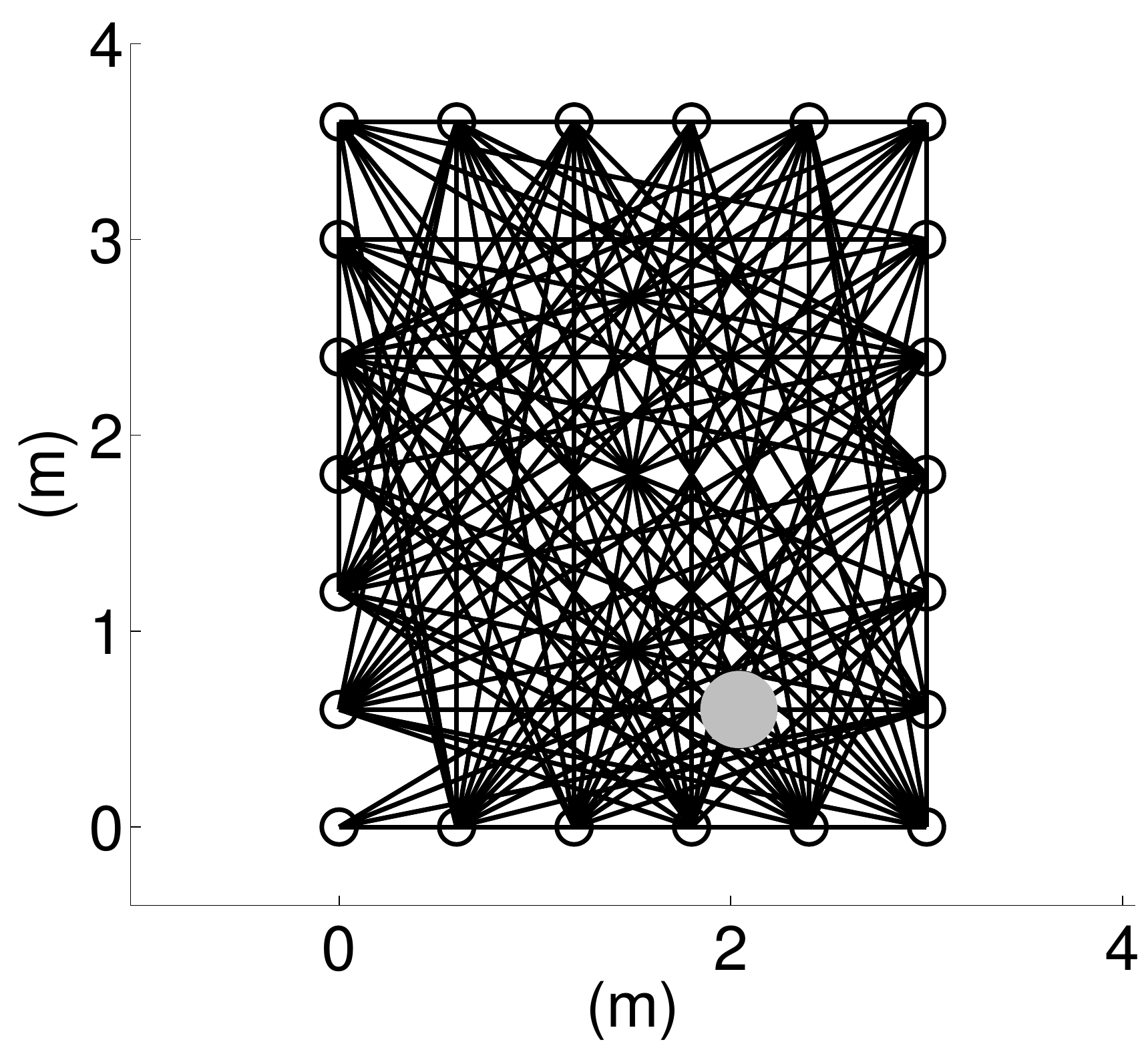}}
\subfigure[FABS-MMCL-O foreground]{
\includegraphics[width=2.57cm]{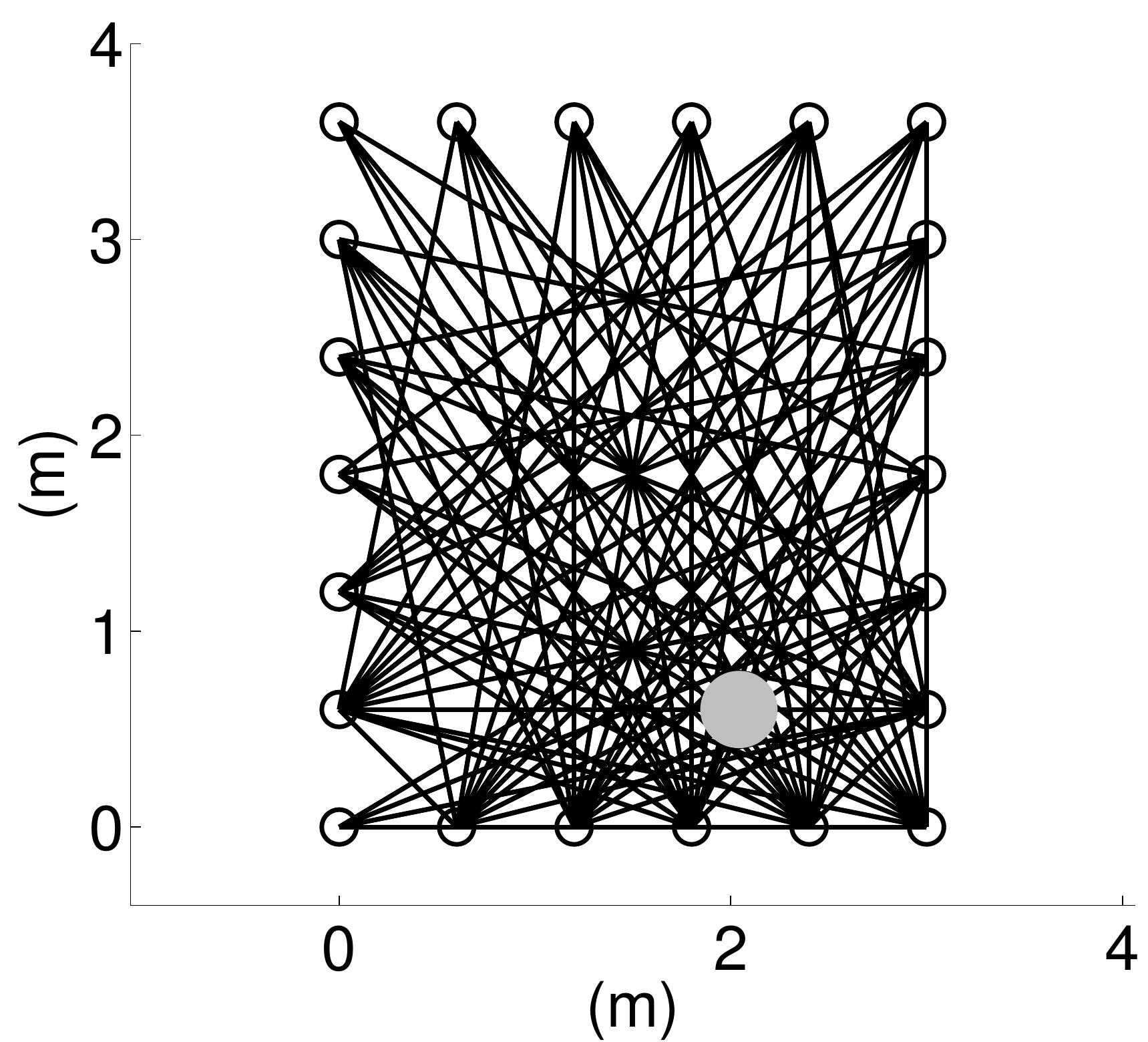}
\label{fig:PlayAgainExperiment2FABSMMCLO}} \\
\subfigure[TBM background]{
\includegraphics[width=2.57cm]{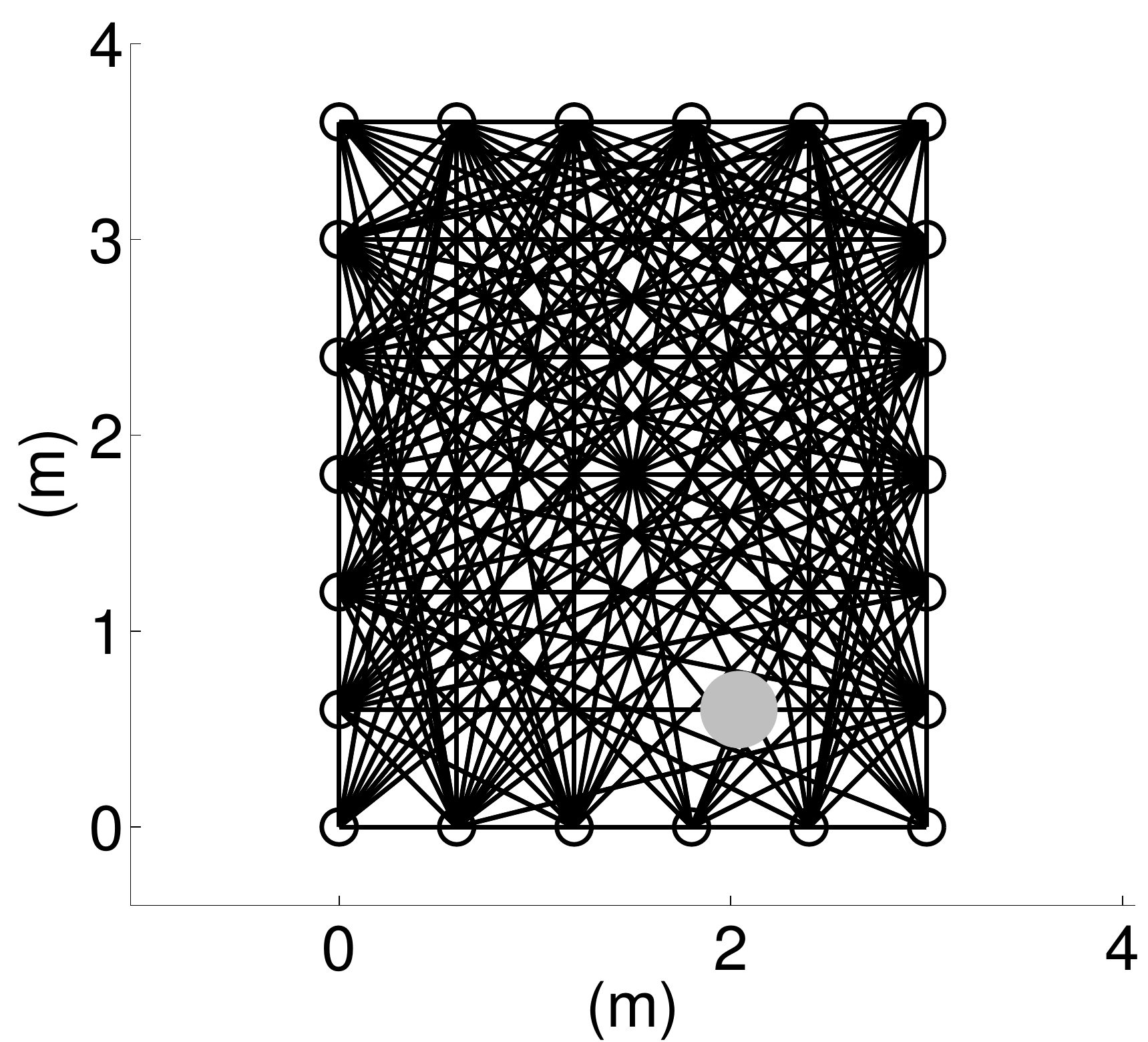}}
\subfigure[FABS background]{
\includegraphics[width=2.57cm]{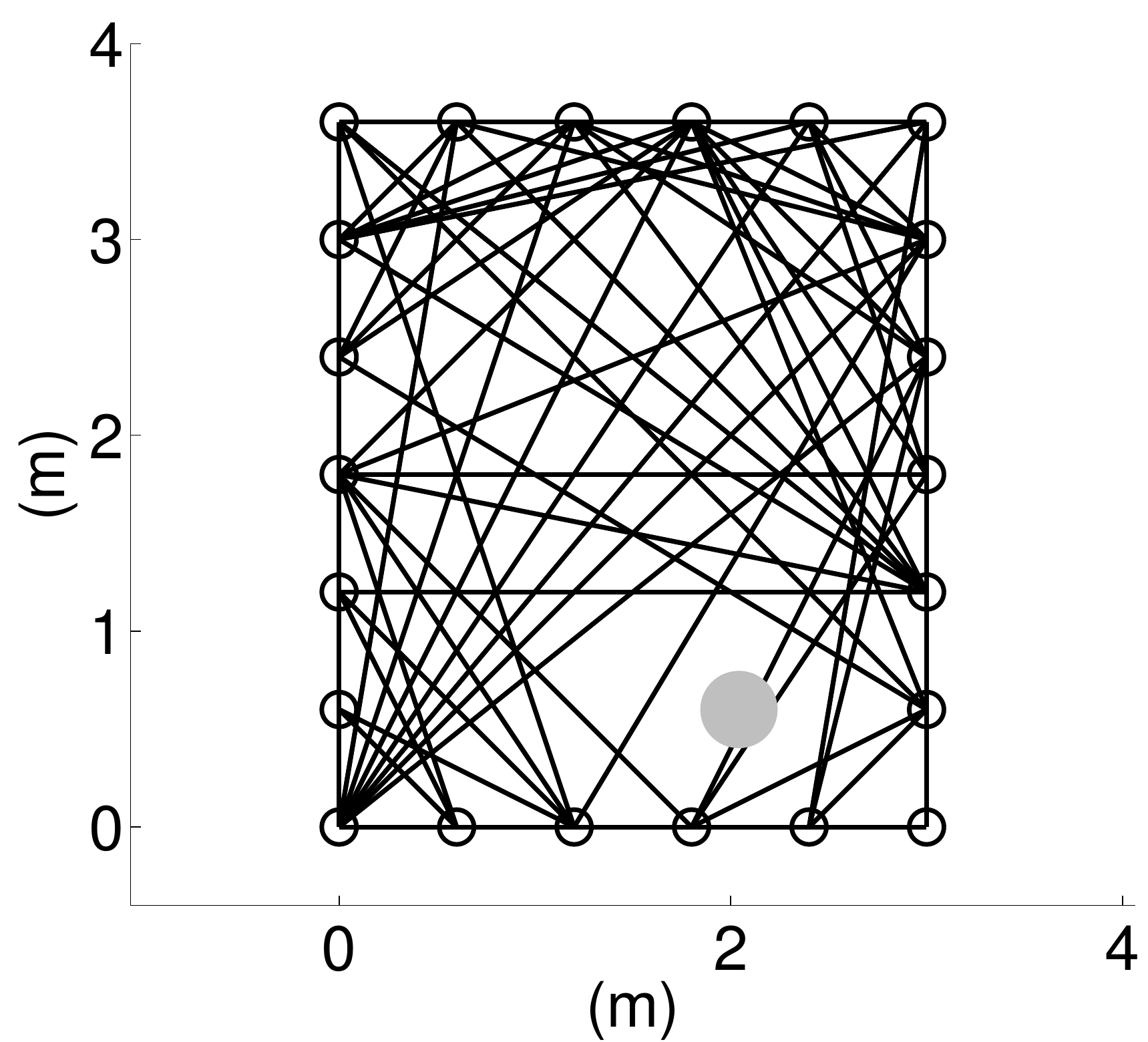}}
\subfigure[FABS-MMCL-O background]{
\includegraphics[width=2.57cm]{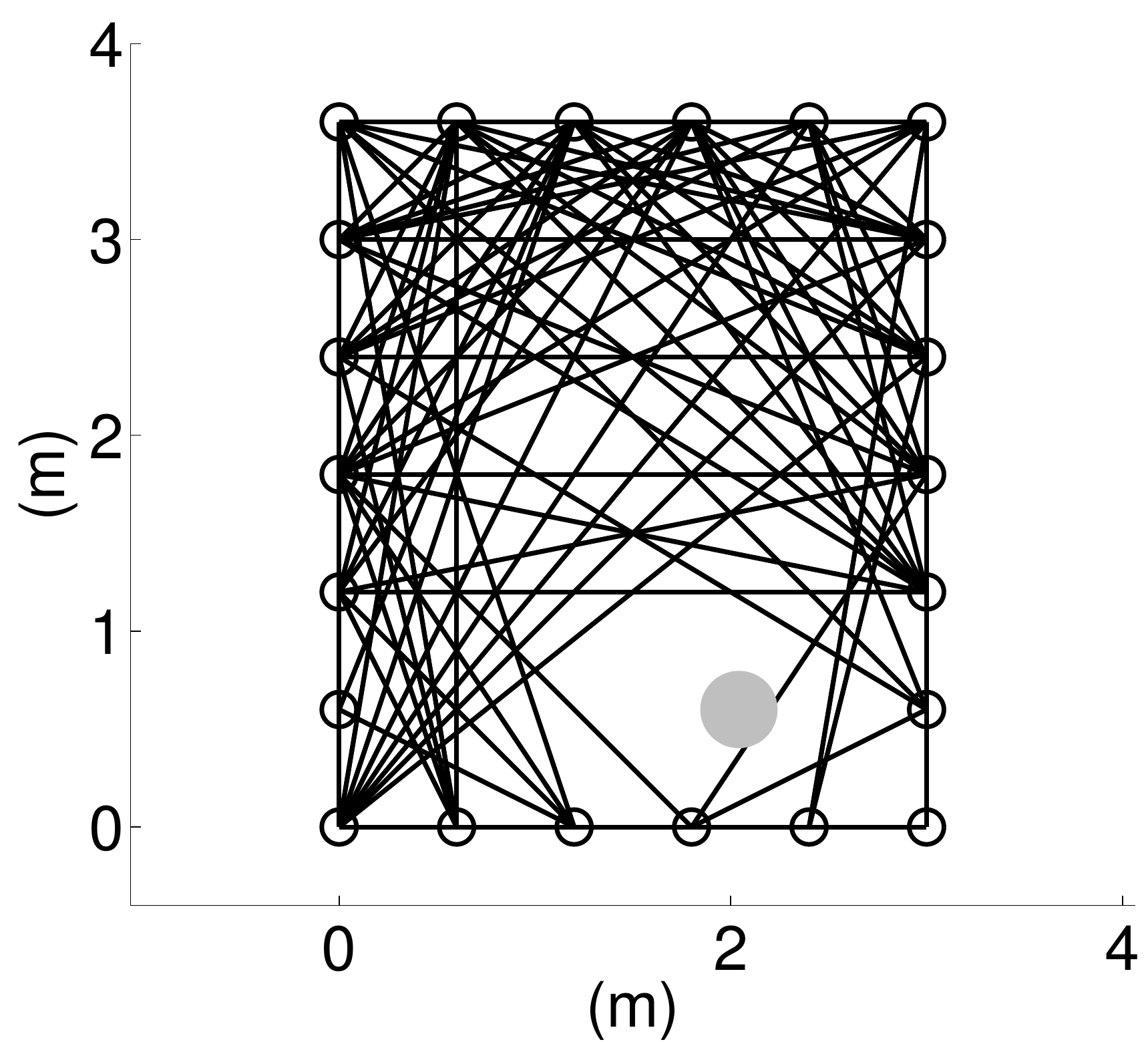}}
\caption{Links assigned to the foreground and background by various background subtraction algorithms for a single {\color{\changedText}representative} frame in experiment set~2.  The known position of the obstruction during this frame is marked.}
\label{fig:PlayAgainExperiment2}
\end{figure}

Once again, we also examine which links are assigned to the foreground and background by each algorithm.  This is presented in Fig.~\ref{fig:PlayAgainExperiment2}.

Unlike with our outdoor experiment set, in this case, we can see that the background subtraction algorithms are behaving as intended.  Namely, TBM selects an initial set of foreground links, FABS expands this set, and FABS-MMCL then prunes the set back.  Although it may seem like we end up with an unintuitively-large proportion of links being assigned to the foreground in Fig.~\ref{fig:PlayAgainExperiment2FABSMMCLO}, we can't argue with the fact that large foreground sets such as this provide the closest approximation to the measured baseline RSS.  Note, however, that this successful approximation is not just a matter of carelessly assigning a large proportion of the network's links to the foreground.  For instance, no matter how we change the parameters of TBM to increase the number of links it assigned to the foreground, we could not decrease the algorithm's estimation error to the level achieved by FABS-MMCL.  The finesse of the more complex approaches is required in order to improve the baseline RSS estimate in this indoor environment.

\subsubsection[Results for Experiment Set 3]{Results for Experiment Set~3}\label{sec:Results3}
\begin{figure}
\centering
\includegraphics[width=3in]{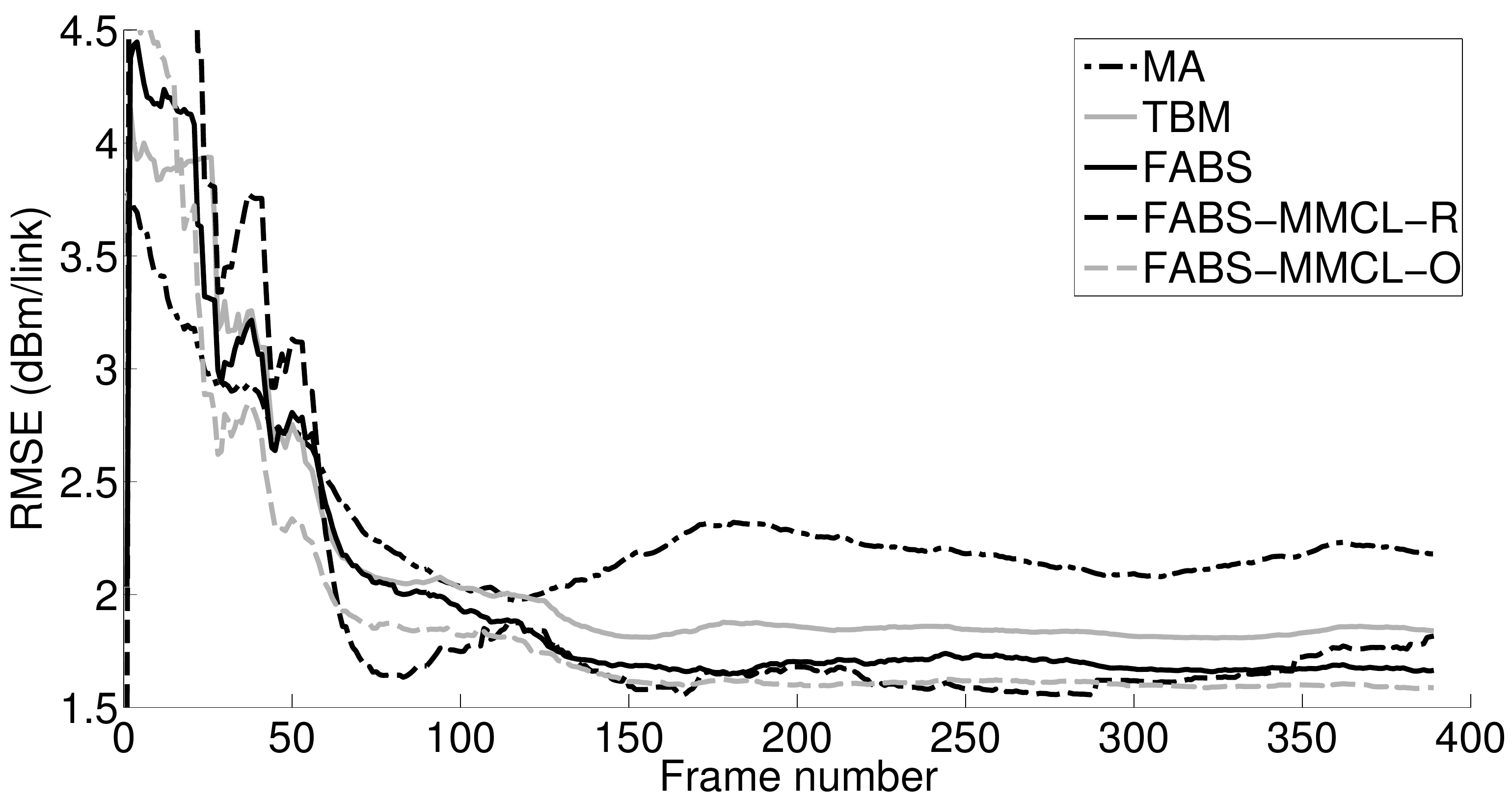}
\caption[Estimation error for experiment set 3]{RMSE obtained when estimating the baseline RSS for experiment set~3.  The~388 frames used were collected while a walker made two round trips around the path in Fig.~\ref{fig:Experiment3SetupDiagram} over a period of~$\approx$~50 seconds.}
\label{fig:Experiment3RMSE}
\end{figure}

Fig.~\ref{fig:Experiment3RMSE} presents the estimation error obtained when using our background subtraction algorithms to estimate the baseline RSS values for experiments in which measurements were collected indoors, through a solid wall.  Table~\ref{tab:Experiment3Parameters} presents the associated parameter values.

\begin{table}
\renewcommand{\arraystretch}{1.3}
\caption[Parameter values used for experiment set 3]{Parameter values used for experiment set~3.}
\label{tab:Experiment3Parameters}
\centering
\begin{tabular}{c||c||c||c||c}
	\hline
	Parameter              & TBM  & FABS & FABS-  & FABS-  \\
	                       &       &      & MMCL-R & MMCL-O \\
  \hline\hline                                         
  $N$                    & 35   & 35    & 35     & 35     \\
  \hline
  $\sigma_t^2$           & 1    & 1     & 1      & 1      \\
  \hline
  $\theta$               & 0.05 & 0.05  & 0.05   & 0.05   \\
  \hline
  $\tau$                 & --   & 1     & 1      & 0      \\
  \hline
  $\sigma_s^2$           & --   & 1     & 5      & 10     \\
  \hline
  $\eta$                 & --   & 5     & 15     & 4      \\
  \hline
  $\gamma$               & --   & --    & 50     & 10     \\
  \hline
  $C$                    & --   & --    & 15     & --     \\
  \hline
\end{tabular}
\end{table}

In this case, we have once again increased the complexity of our environment, which also increases the lowest achievable estimation error seen in Fig.~\ref{fig:Experiment3RMSE}.  Nonetheless, basic TBM still provides an improvement over MA.  Likewise, as in the non-through-wall indoor experiment, we can obtain further improvements over TBM by using FABS and FABS-MMCL-O.

Tracking also becomes more complicated with through-wall measurements.  It now takes the particle filter a long time to converge to the proper track, if indeed it ever does.  For the data we collected, whether we initialize tracking with $\mathbf{\hat{R}_B}$ or with $\mathbf{\tilde{R}_B}$ itself, the algorithm inevitably requires~$\approx$~300 timesteps before the particle filter finds the correct track.  Furthermore, over our~100 realizations initialized with $\mathbf{\tilde{R}_B}$, in~36 cases, the algorithm never converged at all, resulting in very obviously ``lost'' tracks.

In this environment, it thus becomes important to consider not just the tracking error for different $\mathbf{\hat{R}_B}$'s, but also the percentage of lost tracks.  This is summarized in Table~\ref{tab:Experiment3RMSE}.  For instance, when $\mathbf{\hat{R}_B}$ is calculated using MA, every single one of the~100 realizations is a lost track.  Similarly, when $\mathbf{\hat{R}_B}$ is calculated using TBM or with FABS, 95\% and 93\% of the realizations are lost tracks; consequently, the associated tracking error values are not reliable.  It is only the values of $\mathbf{\hat{R}_B}$ produced by FABS-MMCL that allow the tracking algorithm to converge in $>$25\% of realizations.  Then, in these cases, the tracking error is actually consistently lower than the error seen when the tracking algorithm is initialized using $\mathbf{\tilde{R}_B}$.

\begin{table}
\renewcommand{\arraystretch}{1.3}
\caption[Tracking error and lost track percentage for experiment set 3]{Lost track percentage and RMSE for tracking using experiment set~3.}
\label{tab:Experiment3RMSE}
\centering
\begin{tabular}{c||c||c}
	\hline
	 Algorithm & Lost Track \% & Tracking Error (m) \\
	 \hline\hline
	 Calibration Data & 36 & 2.7606$\pm$0.0116 \\
	 \hline
	 MA & 100 & N/A \\
	 \hline
	 TBM & 95 & 1.8364$\pm$0.8932 \\
	 \hline
	 FABS & 93 & 2.6290$\pm$0.0081 \\
	 \hline
	 FABS-MMCL-R & 74 & 1.3068$\pm$0.2080 \\
  \hline
  FABS-MMCL-O & 29 & 1.7380$\pm$0.8578 \\
  \hline
\end{tabular}
\end{table}

\begin{figure}
\centering
\subfigure[TBM foreground]{
\includegraphics[width=2.61cm]{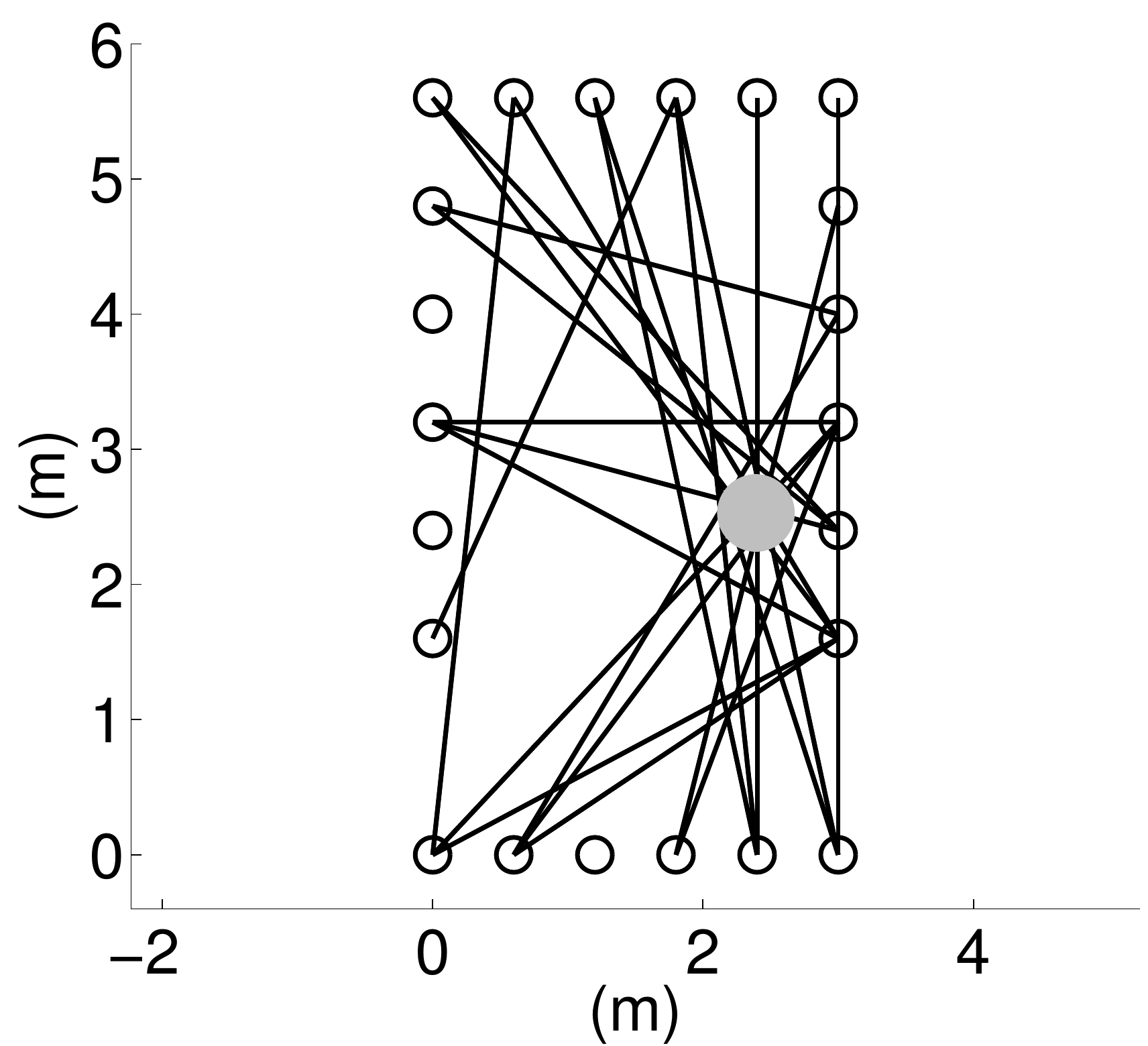}}
\subfigure[FABS foreground]{
\includegraphics[width=2.61cm]{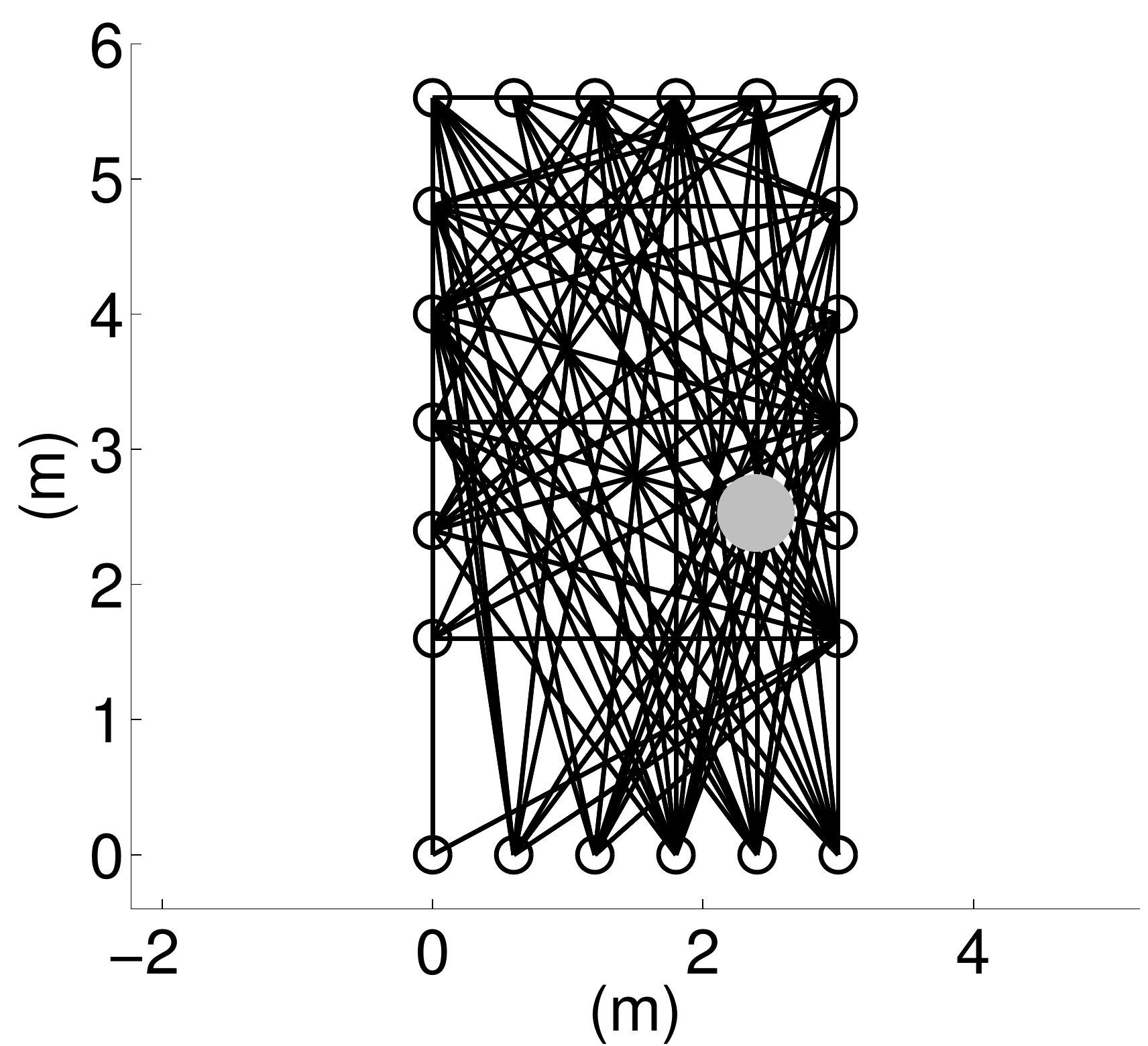}}
\subfigure[FABS-MMCL-O foreground]{
\includegraphics[width=2.61cm]{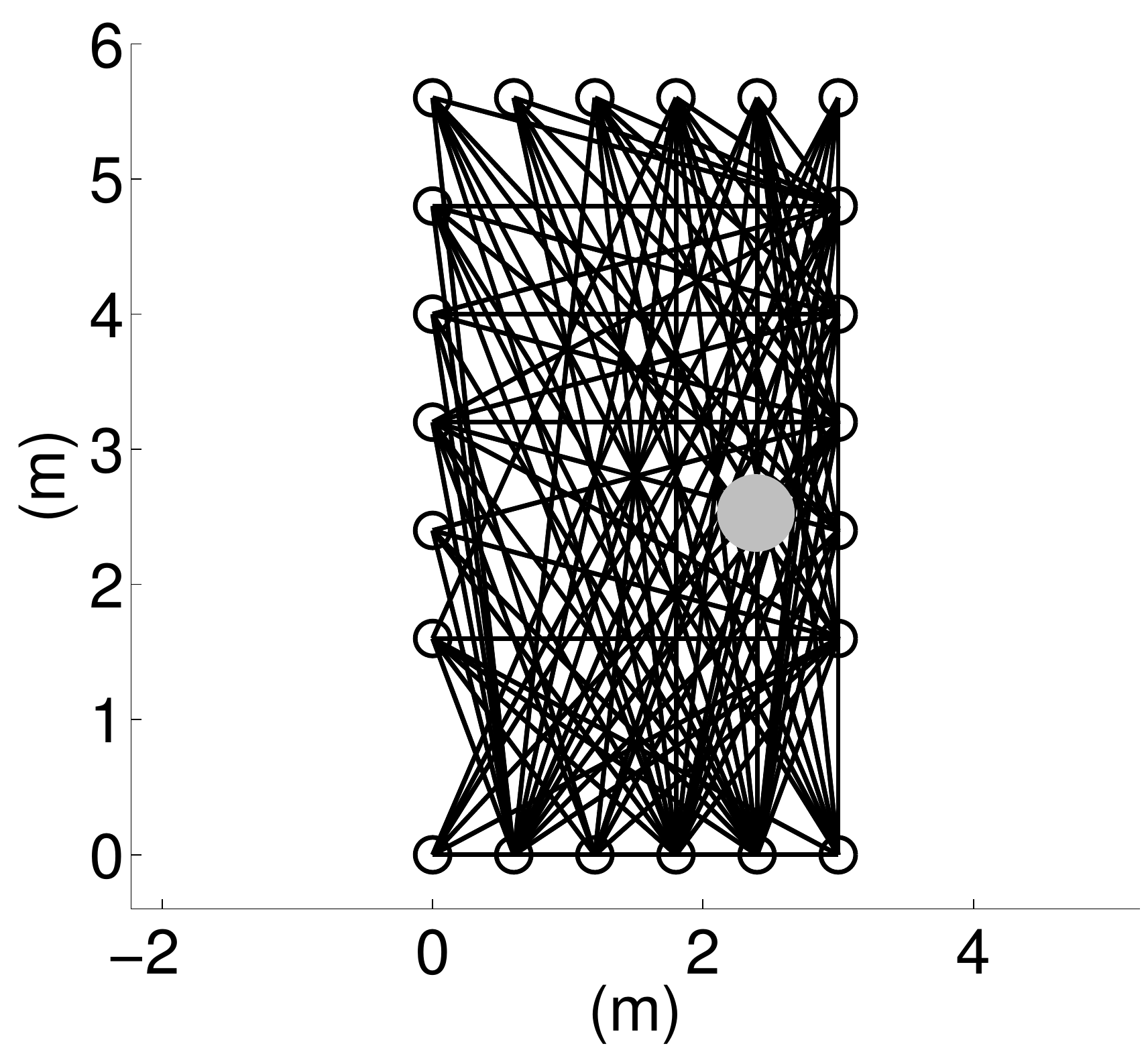}} \\
\subfigure[TBM background]{
\includegraphics[width=2.61cm]{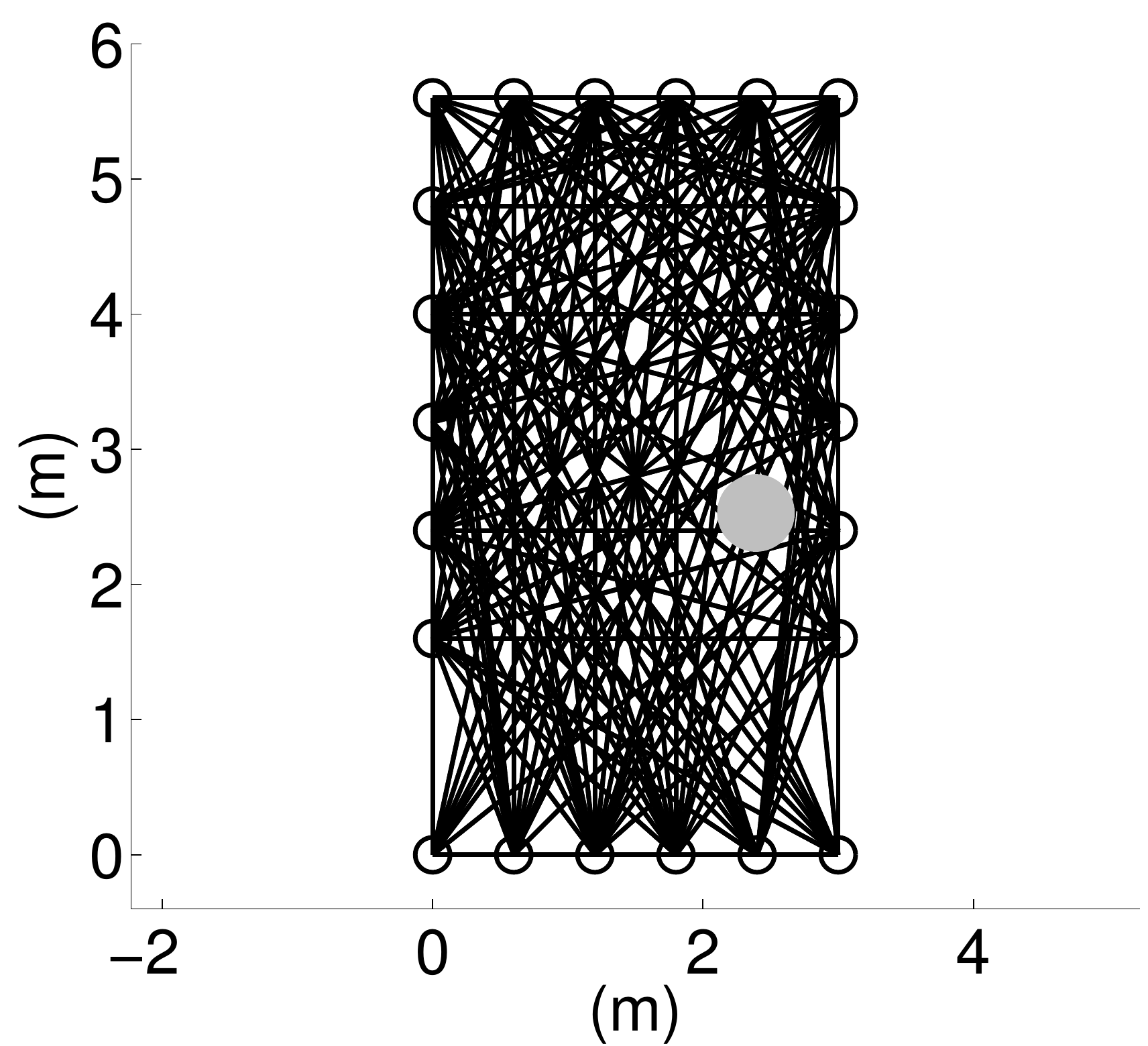}}
\subfigure[FABS background]{
\includegraphics[width=2.61cm]{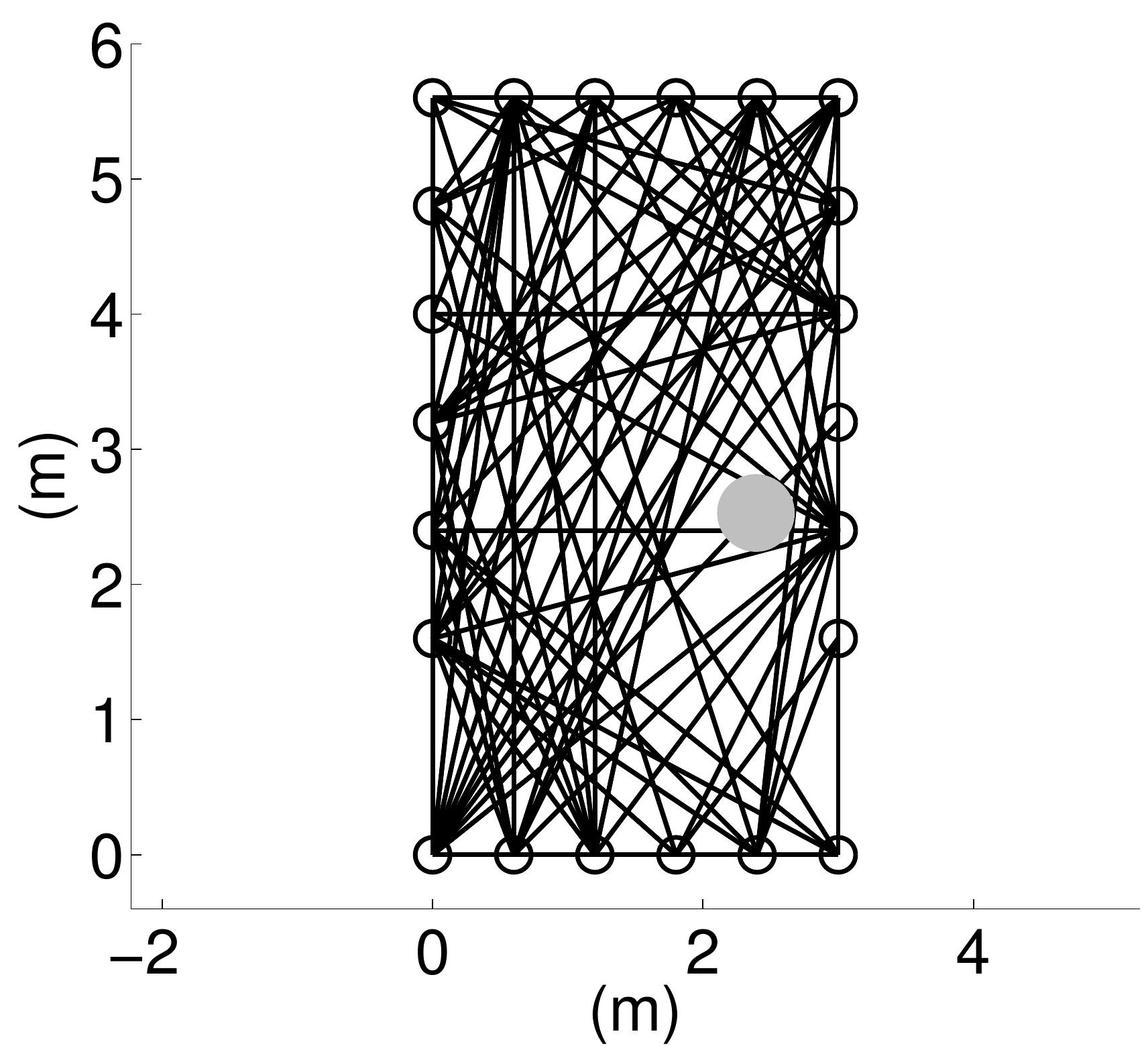}}
\subfigure[FABS-MMCL-O background]{
\includegraphics[width=2.61cm]{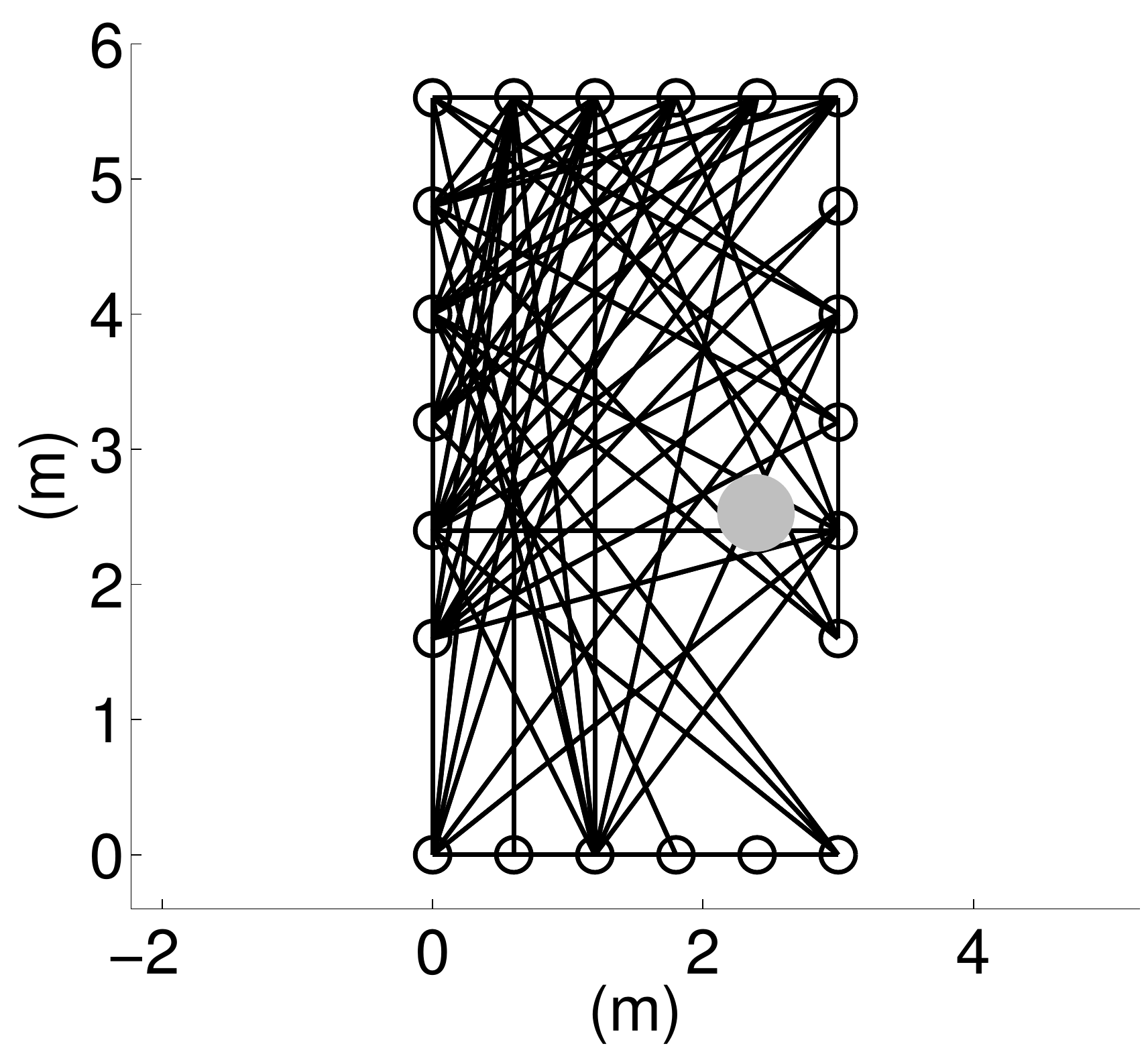}}
\caption{Links selected as being in the foreground and background by various background subtraction algorithms for a single {\color{\changedText}representative} frame in experiment set~3.  The position of the obstruction during this frame is marked.}
\label{fig:PlayAgainExperiment3}
\end{figure}

In Fig.~\ref{fig:PlayAgainExperiment3}, we present a one-frame snapshot of which links in the through-wall environment are being assigned to the foreground and background by each algorithm.  Comparing this to the similar figure presented for the indoor environment, we can see that the layout of the foreground is similar in both cases.

{\color{\changedText}\subsection{Algorithm Speed}\label{sec:Performance}

As previously mentioned, our various background subtraction algorithms differ substantially in terms of the amount of computation they require.  Table~\ref{tab:Performance} presents the average runtime for each algorithm for a single frame (measured over~50 repetitions run on a dual-core~2.13~GHz Intel processor with~2~GB of RAM).  An analysis as to how these runtimes depend on the algorithm parameters can be found in~\cite{Edelstein2011}.

\begin{table}
\renewcommand{\arraystretch}{1.3}
\caption{{\color{\changedText}Observed per-frame run times for each algorithm}}
\label{tab:Performance}
\centering
\begin{tabular}{c||c||c||c||c}
	\hline
	 Exp. & TBM  & FABS & FABS-MMCL- & FABS-MMCL- \\
	 Set & (ms) &  (ms) & R (ms) & O (ms) \\
	 \hline\hline
	 Set 1 & 1.056 & 43.3561 & 240.1544 & -- \\
	 \hline
	 Set 2 & 1.3952 & 23.7115 & 160.0984 & 432.9784 \\
	 \hline
   Set 3 & 1.278 & 62.7504 & 415.7762 & 441.8479 \\
   \hline
\end{tabular}
\end{table}}

\subsection[Sensitivity to Parameter Values]{Sensitivity to Parameter Values}\label{sec:Parameters}

The results reported above show the performance of the proposed background subtraction methods when parameter settings are tuned to yield the best performance. Additional experiments considering the effects of parameter mismatch and investigating sensitivity to parameter values are reported~\cite{Edelstein2011}; they are omitted here due to space limitations. The main conclusion of these experiments is that the proposed methods are most sensitive to changes in the decision thresholds ($\theta$, $\eta$), the natural temperature $\gamma$ used in the Gibbs distributions, and the size of neighborhoods used in FABS-MMCL-R.

\section{Conclusion}\label{chapter:Ch6}

In this work, we explained how several pre-existing background subtraction algorithms could be adapted for use in determining the background, baseline RSS values of links in a WSN while the network is already online and obstructions may already be present.  This constitutes the first set of methods proposed to estimate these values when they cannot be measured during an offline calibration period.  Using experimental data, we have demonstrated that background subtraction techniques can successfully estimate the baseline RSS in a range of different environments.

{\color{\changedText}We can also make recommendations as to which of our proposed algorithms is best-suited to a given environment and application.  Since MA and TBM are both at least moderately effective at estimating the baseline RSS in experiment sets~1 and~2, and since these are the least computationally-complex algorithms---in addition to being the only ones suited for applications such as node localization---these are probably the best choice for non-through-wall applications; if more accurate RSS measurements are needed, e.g., for use with a non-robust RF-sensing application, TBM, FABS or FABS-MMCL may be preferred.  For through-wall applications, the complexity of FABS-MMCL---and in particular of FABS-MMCL-O---will be required.  Thus, all of the methods we proposed fill their own particular niches.}

In addition to examining the performance of our methods, we also obtained a rough estimate of their convergence periods.  This will vary depending on what is occurring inside the region of interest, but in all our experiments, it took less than~300 frames~($\approx$~40 seconds) for our estimates to converge.  Waiting~40 seconds to obtain estimates for the baseline RSS represents a huge improvement over having to wait for the entire network to be evacuated to obtain these values (if evacuating the network is even possible at all).

Finally, we presented some representative runtimes for our different algorithms, and---from our preliminary experiments with different parameter values---we started to gain some insight into how sensitive background subtraction for WSNs is with respect to each algorithm parameter.

\subsection{Future Work}\label{sec:FutureWork}

In the future, we would like to explore more complex ways of modelling the spatial similarity and spatial ergodicity between links, in the hope of seeing better performance in indoor environments.  For instance, we are currently experimenting with creating weighted neighbourhoods which preserve the idea that some of a link's neighbours may be more similar to it (in length, in physical location, in covariance as predicted by the NeSh model~\cite{Patwari2008}, etc.) than others.

There is also work to do to further our understanding of how to optimize the background subtraction parameters.  Although we conducted a brief study of these parameters, a true understanding of their effects will not be complete without an experimental survey which looks at the ideal parameters for different types of environments; for different types, sizes and numbers of obstructions; for different network layouts, etc.

Even in the absence of these improvements and extensions, however, we can still say that using background subtraction to estimate the baseline RSS in wireless sensor networks which are already online is an interesting new technique which has already yielded promising results and which fills a definite need in the domain of RF sensing network applications.

\section*{Acknowledgments}

This research is supported by a grant from the Natural Sciences and Engineering Research Council of Canada. We would like to thank A.~Men, B.~Yang, and Y.~Li at BUPT for providing the experimental data.

\bibliographystyle{abbrv}
\bibliography{Thesis}

\begin{thebibliography}{10}

\bibitem{Agrawal2009}
P.~Agrawal and N.~Patwari.
\newblock Correlated link shadow fading in multi-hop wireless networks.
\newblock {\em IEEE Trans. Wireless Communications}, 8(8):4024--4036, Aug.
  2009.

\bibitem{Besag1986}
J.~Besag.
\newblock On the statistical analysis of dirty pictures.
\newblock {\em Journal of the Royal Statistical Society}, B-48:259--302, 1986.

\bibitem{Edelstein2011}
A.~Edelstein.
\newblock Background subtraction methods for online calibration of baseline
  received signal strength in radio frequency sensing networks.
\newblock Master's thesis, McGill University, Montreal, QC, Canada, Dec. 2011.

\bibitem{Elgammal2002}
A.~Elgammal, R.~Duraiswami, D.~Harwood, and L.~Davis.
\newblock Background and foreground modeling using nonparametric kernel density
  estimation for visual surveillance.
\newblock {\em Proc. IEEE}, 90(7):1151--1163, Jul. 2002.

\bibitem{Gu2009}
Y.~Gu, A.~Lo, and I.~Niemegeers.
\newblock A survey of indoor positioning systems for wireless personal
  networks.
\newblock {\em IEEE Communications Surveys Tutorials}, 11(1):13--32, 2009.

\bibitem{Gudmundson1991}
M.~Gudmundson.
\newblock Correlation model for shadow fading in mobile radio systems.
\newblock {\em Electronics Letters}, 27(23):2145--2146, Nov. 1991.

\bibitem{Li2011}
Y.~Li, X.~Chen, M.~Coates, and B.~Yen.
\newblock Sequential {M}onte {C}arlo radio-frequency tomographic tracking.
\newblock In {\em IEEE Conf. Acoustics, Speech and Signal Processing}, pages
  3976--3979, May 2011.

\bibitem{McHugh2008}
J.~McHugh.
\newblock Probabilistic methods for adaptive background subtraction.
\newblock Master's thesis, Boston University, USA, 2008.

\bibitem{McHugh2009}
J.~McHugh, J.~Konrad, V.~Saligrama, and P.-M. Jodoin.
\newblock Foreground-adaptive background subtraction.
\newblock {\em IEEE Signal Processing Letters}, 16(5):390--393, May 2009.

\bibitem{Moussa2009}
M.~Moussa and M.~Youssef.
\newblock Smart devices for smart environments: device-free passive detection
  in real environments.
\newblock In {\em IEEE Intl. Conf. Pervasive Computing and Communications},
  pages 1--6, Mar. 2009.

\bibitem{Patwari2008}
N.~Patwari and P.~Agrawal.
\newblock Ne{S}h: a joint shadowing model for links in a multi-hop network.
\newblock In {\em IEEE Intl. Conf. Acoustics, Speech and Signal Processing},
  pages 2873--2876, Apr. 2008.

\bibitem{Patwari2005}
N.~Patwari, J.~Ash, S.~Kyperountas, A.~Hero~III, R.~Moses, and N.~Correal.
\newblock Locating the nodes: cooperative localization in wireless sensor
  networks.
\newblock {\em IEEE Signal Processing Magazine}, 22(4):54--69, Jul. 2005.

\bibitem{Patwari2010}
N.~Patwari and J.~Wilson.
\newblock {RF} sensor networks for device-free localization: measurements,
  models, and algorithms.
\newblock {\em Proc. IEEE}, 98(11):1961--1973, Nov. 2010.

\bibitem{Piccardi2004}
M.~Piccardi.
\newblock Background subtraction techniques: a review.
\newblock In {\em IEEE Intl. Conf. Systems, Man and Cybernetics\normalfont{,
  vol. 4}}, pages 3099--3104, Oct. 2004.

\bibitem{Rappaport2002}
T.~Rappaport.
\newblock {\em Wireless Communications Principles and Practice}.
\newblock 2nd ed. Upper Saddle River, NJ: Prentice-Hall, 2002.

\bibitem{CC2530}
{Texas Instruments}.
\newblock {\em A True System-on-Chip Solution for 2.4 GHz IEEE 802.15.4 and
  ZigBee Applications}, 2011.
\newblock available at \url{http://focus.ti.com/lit/ds/symlink/cc2530.pdf}.

\bibitem{Wang2006}
Z.~Wang, E.~Tameh, and A.~Nix.
\newblock Simulating correlated shadowing in mobile multihop relay/ad-hoc
  networks.
\newblock Technical report, IEEE 802.16 Broadband Wireless Access Working
  Group, Jul. 2006.
\newblock available at
  \url{http://www.ieee802.org/16/relay/contrib/C80216j-06_060.pdf}.

\bibitem{Wilson2010Thesis}
J.~Wilson.
\newblock {\em Device-Free Localization with Received Signal Strength
  Measurements in Wireless Networks}.
\newblock PhD thesis, University of Utah, Salt Lake City, UT, USA, Aug. 2010.

\bibitem{Wilson2010}
J.~Wilson and N.~Patwari.
\newblock Radio tomographic imaging with wireless networks.
\newblock {\em IEEE Trans. Mobile Computing}, 9(5):621--632, May 2010.

\bibitem{Wilson2011}
J.~Wilson and N.~Patwari.
\newblock See-through walls: motion tracking using variance-based radio
  tomography networks.
\newblock {\em IEEE Trans. Mobile Computing}, 10(5):612--621, May 2011.

\end{thebibliography}

\end{document}